%% file: fuga.tex
\documentclass[twocolumn]{aa}  

\usepackage{graphicx}
\usepackage{txfonts}
\usepackage[dvipsnames]{xcolor}
\usepackage[pdfencoding=auto,psdextra]{hyperref}
\hypersetup{
    colorlinks=true,
    linkcolor=blue,
    filecolor=magenta,      
    urlcolor=blue,
    citecolor=blue
}

\newcommand{\be}{\begin{equation}}
\newcommand{\ee}{\end{equation}}

\newcommand{\bea}{\begin{eqnarray}}
\newcommand{\eea}{\end{eqnarray}}

\newcommand{\bem}{\begin{multline}}
\newcommand{\eem}{\end{multline}}

\newcommand{\nn}{\nonumber\\}
\newcommand{\half}{{\textstyle\frac12}}

\newcommand{\ve}[1]{\mathbf{#1}}
\newcommand{\fuga}{\texttt{FuGa3D}}
\newcommand{\realspace}{\texttt{Realspace}}
\newcommand{\corrfunc}{\texttt{Corrfunc}}
\newcommand{\healpix}{\texttt{Healpix}}
\newcommand{\Euclid}{{Euclid}}
\newcommand{\lmax}{\ensuremath{\ell_\text{max}}}
\newcommand{\nside}{\ensuremath{N_\text{side}}}
\newcommand{\zdelta}{\ensuremath{z_\Delta}}
\newcommand{\tnside}{\texttt{nside\_high}}
\newcommand{\tnbase}{\texttt{nside\_base}}
\newcommand{\tzdelta}{\texttt{z\_delta}}

\usepackage{natbib}
\bibpunct{(}{)}{;}{a}{}{,}

\begin{document}

\title{FuGa3D: Fast full-sky analysis of Galaxy catalogs in 3D}

\author{Elina Keih\"anen\inst{1,2}, Jani Haapala\inst{1}, Valtteri Lindholm\inst{1,2}, Martin Reinecke\inst{3}, Susan Rissanen\inst{1}, Jussi V\"aliviita\inst{1,2}, Akke Viitanen\inst{1,4}}
\authorrunning{Keih\"anen et al.}

\institute{
University of Helsinki, Department of Physics,
P.O.~Box 64, FIN-00014, Helsinki, Finland
\and
Helsinki Institute of Physics,
P.O.~Box 64, FIN-00014, Helsinki, Finland
\and
Max-Planck-Institut f\"ur Astrophysik, Karl-Schwarzschild-Str.~1, 85741 Garching, Germany
\and
{Department of Astronomy, University of Geneva, ch. d’Ecogia 16,
1290 Versoix, Switzerland}}

\abstract{We present \fuga, a code for fast computation of correlation functions and power spectra for galaxy survey observables, including galaxy clustering and cosmic shear. We define the redshift-space correlation function (RCF) as the correlation function defined in the parameter space of two redshifts and an angular separation angle. Assuming that there is no preferred direction in the sky, these parameters fully define the relative position of two galaxies, independently of the assumed cosmological model.
Once the RCF is constructed, it is easy to compute derived correlation metrics, such as the real-space clustering correlation function and its multipoles. We further define the redshift-space power spectrum as the harmonic counterpart of the RCF, and show that it can be computed efficiently using the discrete galaxy coordinates.  We validate the code with simulated mock catalogs.
Computing the RCF and the two-point correlation function at 1.5 Mpc (3 Mpc) resolution for a MICE simulation with 46 million galaxies, took 47 node-min for clustering only, and 7.3 node-hours with shear analysis included.
}

\maketitle

\section{Introduction}

Understanding the large-scale structure and the expansion history of the universe is a key goal of modern cosmology. 
Two powerful observational techniques, galaxy clustering and weak gravitational lensing, provide complementary probes of the distribution of matter in our universe. Galaxy clustering (GC) measures the statistical distribution of galaxies. Weak lensing (WL) measures the subtle distortion of distant galaxy shapes due to gravitational deflection by intervening matter and offers a direct probe of the total mass distribution, independent of galaxy bias. Together, these methods allow for precise tests of cosmological models and constraints on the nature of dark energy. 

The largest galaxy clustering surveys before the 2020s are provided by 2dFGRS \citep{Colless:2003wz} and SDSS I-IV \citep{SDSS:2000hjo,SDSS:2008tqn} and its various subprojects.
The most recent observations come from DESI  \citep{DESI:2025fxa}, and hint towards evolving 
dark energy.
KiDS \citep{Heymans:2020gsg} performs a WL survey,
and DES \citep{DES:2016jjg} and HSC \citep{Li:2023tui} combine clustering statistics with a WL survey.

The spectroscopic survey of the \Euclid\ mission \citep{EuclidOverview} will measure the locations of tens of million of galaxies at unprecedented accuracy. The photometric survey will image the WL effect on the shapes of over a billion galaxies.
The LSST survey of Vera C. Rubin Observatory \citep{LSST:2008ijt} is expected to provide a database of 20 billion galaxies.
The Nancy Grace Roman Space Telescope \citep{Akeson:2019biv} is scheduled for launch in 2027.
The analysis of these ever-increasing data sets calls for efficient analysis methods.

One of the fundamental tools used to quantify the clustering of galaxies is the two-point correlation function (2PCF) \citep{Peebles:1980yev}, which measures the excess probability of finding galaxy pairs at a given separation compared to a random distribution.
Besides the simplest isotropic correlation function $\xi(r)$, which is a function of the distance modulus $r$, one can define anisotropic variants of the 2PCF, to capture redshift-space distortion (RSD) effects that arise from the proper motion of the galaxies or coherent flow of the cosmic fluid \citep{Kaiser:1987qv}. 

Constructing the real-space 2PCF requires knowledge of the geometry and the expansion history of the universe.  The radial distance to a galaxy cannot be measured directly, but must be inferred from the observed redshift.  To convert the redshift into a distance, one must assume a  {\em fiducial} cosmological model and assign values to the cosmological parameters.
Within the cosmological standard model ($\Lambda$CDM), the matter density parameter $\Omega_\mathrm{m}$, vacuum energy density parameter $\Omega_\Lambda$, and expansion rate $H_0$ in particular affect the redshift-distance relation.  
If the fiducial model differs from the true cosmology, the resulting 2PCF is distorted \citep{ 
Alcock:1979mp,Ballinger:1996cd}
It is possible to iterate the process with updated parameter values, but for modern large galaxy surveys, re-evaluating the 2PCF is a computationally heavy task.  Standard methods to estimate the 2PCF, like the Landy-Szalay \citep{Landy:1993yu} estimator, rely on counting galaxy pairs, a task whose computational complexity scales as proportional to the square of the number of galaxies, with survey volume fixed. Alternatively, if the distortion is small, one can apply a correction to the 2PCF, to compensate for the difference between the fiducial model and the derived one. 
%To see how this was done for the BOSS Survey for example, see \cite{BOSS:2016apd,BOSS:2016wmc}.
Literature on RSD analyses is extensive. To mention a few, see \cite{Peacock:2001gs,Okumura:2007br, Guzzo:2008ac,Beutler:2012px,BOSS:2012xge,eBOSS:2018whb,eBOSS:2018cab,eBOSS:2020lta}.

Time evolution further complicates the analysis of the galaxy distribution. In deep surveys, the conventional approach is to divide the survey into tomographic redshift shells. The bins should be narrow enough so that the time evolution effects within a shell can be ignored, but wide enough to include all scales of interest, which are somewhat contradictory requirements. Should one want to switch from one tomographic binning scheme to another, the clustering 2PCF must be recomputed, which again is a computationally heavy process.

Weak lensing is typically quantified in terms of angular correlation functions or angular spectra in tomographic redshift bins, which do not require converting redshifts to distances (e.g. \citealt{Schneider:2002jd,Kilbinger:2014cea}).
WL analysis does not depend on the choice of a fiducial model in the same way as clustering analysis. WL studies could still benefit from a way of re-binning the data to another tomographic binning without full recomputation.

The aim of this paper is twofold:

1. We want to describe the correlation structure in the observed galaxy distribution in terms of directly observable quantities, without assuming a cosmological model. In particular, we use the redshift $z$ as the radial coordinate. For this purpose, we define the redshift-space correlation function (RCF), and its harmonic counterpart, the redshift-space power spectrum.

2. We introduce \fuga, a code for fast analysis of galaxy catalogs.
\fuga\ -- {\bf Fu}ll-sky analysis of {\bf Ga}laxy catalogs in {\bf 3D} -- computes redshift-space correlation functions and power spectra for galaxy clustering, weak lensing shear, and cross-correlation between the two. A number of auxiliary tools allow the primary data objects to be converted into the conventional real-space correlation function, or to re-bin them to an angular correlation function or angular power spectrum in an arbitrary redshift range.  

The paper is structured as follows: in Section 2 we introduce the redshift-space correlation function and power spectrum, and describe the theoretical framework. In Section 3 we introduce the \fuga\ code and describe the computation of the redshift-space correlation function.  In Section 4 we extend the analysis to harmonic space, and show that the power spectra can be evaluated efficiently using discrete galaxy coordinates. In Section 5 we validate the code using simulated galaxy mocks and demonstrate the usefulness of the RCF.  We conclude and discuss future developments in Section 6.

%%%%%%%%%%%%%%%%%%%%%%%%%%%%%%
\section{Correlation function in redshift space}

\subsection{Clustering correlation function}

The conventional galaxy 2PCF is defined as the excess probability of finding two galaxies in volume elements  $dV_1,dV_2$ at a separation distance of $\ve r$, as compared to an unclustered distribution,
\be
dP(\ve r_0;\ve r_0+\ve r)  = \bar n_1\bar n_2 (1+\xi(\ve r)) dV_1dV_2,
\label{eq:genCorrDef}
\ee
where $\bar n_1,\bar n_2$ is the average density of observable galaxies at the two locations, and $\xi(\ve r)$ defines the correlation function. The expression assumes statistical homogeneity; the correlation function depends only on the relative distance $\ve r$.

In a statistically homogeneous and isotropic universe, $\xi(\ve r)$ would be a function of the modulus of $\ve r$. We could ignore the orientation of the galaxy pair, and write $\xi(\ve r)=\xi(r)$. Redshift-space distortions (RSD) complicate the picture.
The anisotropy of the correlation structure is captured by the two-dimensional correlation function $\xi(r,\mu)$, where $\mu$ is the cosine of the angle between the local line-of-sight (LOS) and the line connecting two galaxies.
It is convenient to expand this in Legendre polynomials as
\be
\xi(r,\mu) =\sum_\ell \xi_\ell(r)L_\ell(\mu) \,.
\label{eq:multipoleDef}
\ee
This defines the correlation function multipoles $\xi_\ell(r)$. In the absence of anisotropy, only the monopole $\xi_0(r)$ differs from zero.
The expansion is not unique, but depends on the definition of the local LOS.

Even the correlation function of Eq.~(\ref{eq:genCorrDef}) is not fully general. It assumes statistical homogeneity, that the density of objects and the correlation function are equal everywhere in space. While this may be true at a fixed cosmological time, it does not hold for an observed light-cone, which is a combination of galaxies at different distances and thus at different stages of their evolution. The correlation function evolves with time, making the observed correlation a function of distance.

We now define the {\em redshift-space correlation function} (RCF) $\xi(z_1,z_2,\theta)$ for clustering as the excess probability of finding two galaxies at redshifts $z_1$ and $z_2$, and separated by angular distance $\theta$.
Let $\Omega$ denote a direction on the sky, specified by the spherical coordinates $\{\vartheta,\varphi\}$, or by RA and DEC.
The position of a galaxy in redshift space is fully defined by the coordinates $\{z,\Omega\}$.
The probability of finding two galaxies at positions $\{z_1,\Omega_1\}$, $\{z_2,\Omega_2\}$ can be written as
\be
dP(z_1,\Omega_1;z_2,\Omega_2) 
= \bar n_1\bar n_2 (1+\xi(z_1,z_2,\theta))dV_1dV_2,
\label{eq:RCFdef1}
\ee
where $\theta$
is the angular distance between directions $\Omega_1,\Omega_2$, and
$\bar n,\bar n_2$ represents the unclustered density. We have written it separately for locations 1 and 2, to accommodate a distance-dependent selection function (the fainter the object, the closer it must be to be observable), or position-dependent selection effects.

Throughout this work we assume angular isotropy, that is, the correlation depends on the angular distance between the locations, but remains unchanged if the pair is rotated to another orientation, or transported to another position on the sky. In other words, there is no preferred direction or orientation on the sky.
In redshift, we cannot do a similar reduction of variables. The same separation in redshift corresponds to a different physical distance depending on the redshift; thus we cannot reduce the redshift dimension without assuming a cosmological model, which is exactly what we want to avoid. The redshift-space correlation function is necessarily a function of two redshifts.

Under the assumption of angular isotropy, the RCF contains all information on the correlation structure of the galaxy sample, in a form that is independent of the assumed cosmological model, or time evolution. It is also independent of the definition of the LOS. The RCF can be regarded simply as a tomographic angular correlation function, for infinitesimally thin redshift shells.

Constructing $\xi(z_1,z_2,\theta)$ from a large galaxy catalog is a computationally heavy operation (just as constructing the conventional 2PCF is), but once constructed, computing more conventional correlation measures from it is a very fast process. To mention a few possibilities, we can \\
- integrate over $z_1,z_2$ to obtain the angular correlation function for an arbitrary redshift range. \\
- adopt a cosmological model to convert redshifts to physical distances and to rebin the correlation function to the real-space 2PCF.

Alternatively, the clustering correlation function can be defined through the relative density fluctuation
\be
\delta(z,\Omega) = \frac{n(z,\Omega)}{\langle n(z,\Omega)\rangle} -1   \label{eq:deltaDef}
\ee
where $n(z,\Omega)$ is the observed galaxy density, and $\langle n(z,\Omega)\rangle$ represents the mean density at the same location. We write it as a function of the spatial coordinates $z,\Omega$, to allow for location-dependent selection effects. The brackets denote an ensemble average. The local galaxy density is thought to represent a random deviation around the mean value.
By definition $\langle\delta\rangle=0$.
The RCF is then defined as
\be
\xi(z_1,z_2,\theta) = \langle\delta(z_1,\Omega_1) \delta(z_2,\Omega_2) \rangle. \label{eq:RCFdef2}
\ee
With $\bar n$ identified with $\langle n\rangle$, the definition of Eq.~(\ref{eq:RCFdef2}) is equivalent to that of Eq.~(\ref{eq:RCFdef1}). When viewed at very fine resolution, the galaxy density $n$ becomes a collection of delta peaks. Consequently, $\delta$ becomes a function that takes a negative value everywhere except at the locations of observed galaxies, where it is infinite.
This makes the definition of Eq.~(\ref{eq:RCFdef1}) slightly more intuitive, but mathematically the two definitions are equivalent.

\subsection{Shear correlation function}

In a similar manner we define the RCF for the shear field $\gamma$, as a function of two redshifts and an angular separation,
as
\bea
\xi_{\rm tt}(z_1,z_2,\theta) &=& \langle\gamma_{\rm t}(z_1,\Omega_1)\gamma_{\rm t}(z_2,\Omega_2)\rangle \nn
\xi_{\times\times}(z_1,z_2,\theta) &=& \langle\gamma_\times(z_1,\Omega_1)\gamma_\times(z_2,\Omega_2)\rangle 
\eea
where $\gamma_{\rm t},\gamma_\times$ are the tangential and cross shear components, defined with respect to the great circle connecting $\Omega_1$ and $\Omega_2$.

It is customary to combine the correlation function components into ``plus'' and ``minus'' components 
\bea
\xi_+(z_1,z_2,\theta)&=&\xi_{\rm tt}(z_1,z_2,\theta)+\xi_{\times\times}(z_1,z_2,\theta) \nonumber \\
\xi_-(z_1,z_2,\theta) &=& \xi_{\rm tt}(z_1,z_2,\theta)-\xi_{\times\times}(z_1,z_2,\theta).
\eea
In the same way, one can define cross-correlation components $\xi_{\rm t\times},\xi_{\times\rm t}$, however, these are usually expected to vanish on grounds of parity symmetry.

To define the cross-correlation function between density and shear, we make use of Eq.~(\ref{eq:deltaDef}). The cross-correlation between density and tangential shear becomes
\be
\xi_{\rm ct}(z_1,z_2,\theta) = \langle \delta(z_1,\Omega_1)\gamma_{\rm t}(z_2,\Omega_2)\rangle.
\ee
Again, based on parity symmetry, the component $\xi_{\rm c\times}$ is expected to vanish.

The following symmetry relation should be obvious:
\bea
\xi_{\rm cc}(z_1,z_2,\theta) &=& \xi_{\rm cc}(z_2,z_1,\theta) \nn
\xi_+(z_1,z_2,\theta) &=&  \xi_+(z_2,z_1,\theta) \nn
\xi_-(z_1,z_2,\theta) &=&  \xi_-(z_2,z_1,\theta) \nn
\xi_{\rm ct}(z_1,z_2,\theta) &=& \xi_{\rm tc}(z_2,z_1,\theta) \,.
\eea
Similar relations hold for other auto-correlation and cross-correlation components. Here we used the subscript cc to denote the clustering correlation.

\subsection{Redshift-space power spectrum}
\label{sec:powspectheory}

We further define the redshift-space (angular) power spectra (RP) $P^{XY}_\ell(z_1,z_2)$, where $X,Y=C,E,B$, as the harmonic-space counterparts of the RCF components. Here, $C$ refers to the density fluctuation. The $T,\times$ components of the shear field have a correspondence in the $E,B$ components in harmonic space. 
The relations between the non-vanishing components of RCF and RP are given by 
\bea
&&\xi_{\rm cc}(z_1,z_2,\theta) = \sum_\ell \frac{2\ell+1}{4\pi} C^{CC}_\ell(z_1,z_2) d^\ell_{00}(\theta) \nn
&&\xi_+(z_1,z_2,\theta) = \sum_\ell \frac{2\ell+1}{4\pi} (C^{EE}_\ell(z_1,z_2) +C^{BB}_\ell(z_1,z_2)) d^\ell_{2,2}(\theta) \nn
&&\xi_-(z_1,z_2,\theta) = \sum_\ell \frac{2\ell+1}{4\pi} (C^{EE}_\ell(z_1,z_2) -C^{BB}_\ell(z_1,z_2)) d^\ell_{2,-2}(\theta) \nn
&&\xi_{\rm ct}(z_1,z_2,\theta) = -\sum_\ell \frac{2\ell+1}{4\pi} C^{CE}_\ell(z_1,z_2)  d^\ell_{20}(\theta)
\label{eq:powspecToCorr}
\eea
%
%&&\xi_{\rm c\times}(z_1,z_2,\theta) = -\sum_\ell \frac{2\ell+1}{4\pi} C^{CB}_\ell(z_1,z_2)  d^\ell_{20}(\theta) \nn
%&&\xi_{\rm t\times}(z_1,z_2,\theta) +\xi_{\times\rm t}(z_1,z_2,\theta) = \nn 
%&&\qquad\sum_\ell \frac{2\ell+1}{4\pi} (C^{EB}_\ell(z_1,z_2)+C^{BE}_\ell(z_1,z_2) )  d^\ell_{2,-2}(\theta) \nn
%&&\xi_{\rm t\times}(z_1,z_2,\theta) -\xi_{\times\rm t}(z_1,z_2,\theta) = \nn 
%&&\qquad\sum_\ell \frac{2\ell+1}{4\pi} (C^{EB}_\ell(z_1,z_2)-C^{BE}_\ell(z_1,z_2) )  d^\ell_{2,2}(\theta) 
%
where $d^\ell_{ss'}(\theta)$ are the reduced Wigner functions (see, e.g., \citealt{Varshalovich:1988ifq}).
The derivation of these results, including the zero components, is given in Appendix \ref{appendix:xi}.

%%%%%%%%%%%%%%%%%%
\section{Computation on sphere}

\subsection{\fuga}

In this section we describe the \fuga\ analysis code.
\fuga\ takes as input a galaxy catalog. The catalog must include, for each galaxy object, its location on the celestial sphere (RA, DEC), redshift,
and, optionally, the shear components $\gamma_1,\gamma_2$.
From these inputs, \fuga\ computes a number of output products.
The primary outputs are:
the redshift-space correlation function $\xi_X(z_1,z_2,\theta)$, the integrated angular correlation function $\xi_X(\theta)$ for the full redshift range,
angular power spectrum $P_{Y\ell}$, and redshift-space power spectrum $P_{Y\ell}(z_1,z_2)$.
The subscripts $X,Y$ refer to the components related to galaxy density and shear.

Further derived data products can be constructed with auxiliary tools, as will be described later in this paper: The real-space 2PCF of galaxy density is constructed from the RCF with the \realspace\ tool, given a cosmological model.
Another tool corrects the raw power spectrum for the effects of the sky mask.

%%%%%%%%%%%%%%%%%%
\subsection{Resolution and storage}
\label{sec:resolution}

\fuga\ operates in a discretized space.
The computations are carried out on a 3D grid which combines uniformly spaced redshift shells with \healpix\ \citep{Gorski:2004by} pixels. The pixel resolution is set by parameter \tnside, and the redshift resolution by parameter \zdelta, both user-defined parameters.  The redshift resolution of the grid also determines the redshift resolution of the output RCF.

\fuga\ makes use of a hierarchical pixelization scheme which involves two distinct resolutions: 
The sky is first divided into low-resolution pixels, at a resolution set by parameter \texttt{nside\_base}.
The default value \texttt{nside\_base=32} divides the sky to 12\,288 pixels.
We refer to the low-resolution pixels as {\em base pixels}.
The base pixels serve several purposes, including bookkeeping and handling the radial selection function. 
Each low-resolution pixel is divided into smaller pixels at \tnside\ resolution. 
The actual computations are carried out at this high resolution. Moving back and forth between the two resolutions is straightforward in the \texttt{NESTED} pixel ordering scheme of \healpix.

The computation of the galaxy correlation function requires knowledge of the selection function $S(z,\Omega)$.
The selection function tells which fraction of all galaxies is observable at a given location.
In our implementation, the selection function is assumed to be isotropic within one base pixel, that is, a function of redshift only.
This assumption is crucial for the speed-up algorithm implemented in \fuga, as we will explain shortly.
The selection function can either be estimated from the data catalog itself or be provided in the form of an external random catalog.

\begin{figure}
\centering
\includegraphics[width=7cm]{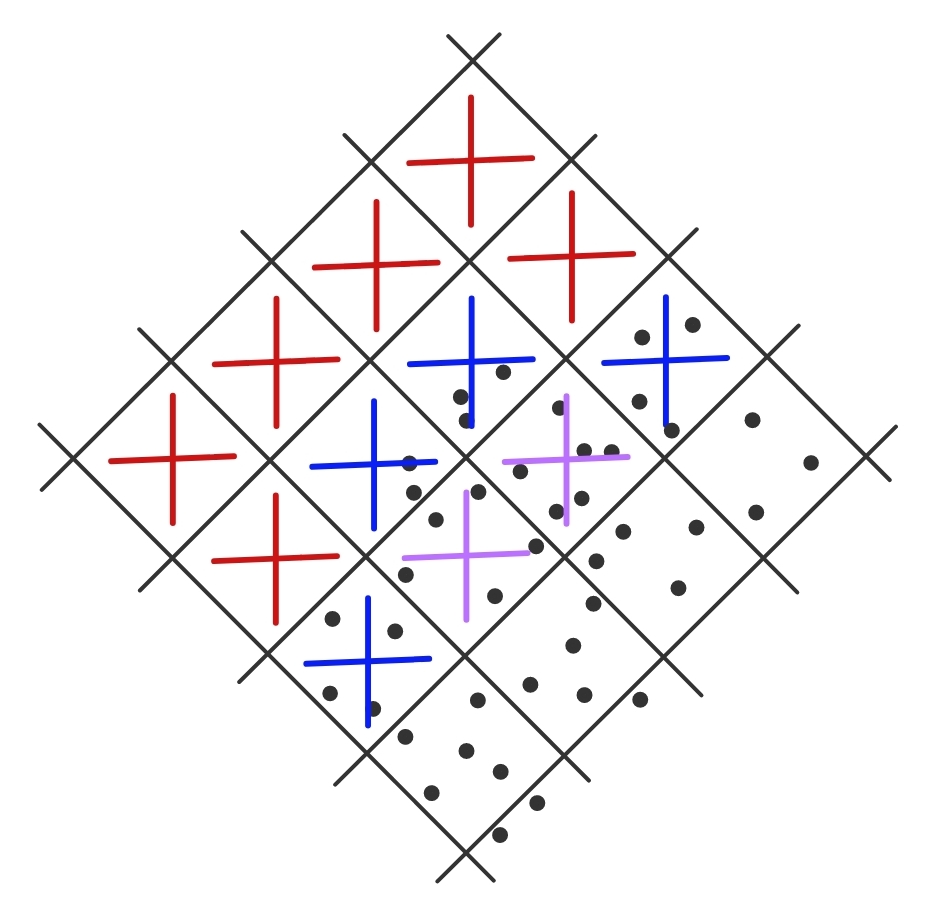}
\caption{Partially filled base pixels are discarded from analysis, to ensure uniform coverage over the full survey area.
Empty base pixels are marked by red crosses.
Pixels with an empty pixel as neighbor, either directly (blue) or diagonally (purple), are interpreted as being on the survey boundary.
Both the empty pixels and their neighbors are discarded from further analysis.
\label{fig:boundary}}
\end{figure}

From the requirement of the isotropic selection function follows that the survey boundary is not allowed to pass through a base pixel.
In the first processing step, the data catalog is trimmed at the survey boundaries to match the low-resolution pixelization.
Partially filled pixels are removed from further processing. This is achieved through the following procedure.
We examine the distribution of galaxies per base pixel, ignoring the redshift.
We identify all low-resolution pixels that have empty pixels as neighbors (Fig.~\ref{fig:boundary}).
We interpret these pixels as being partially filled, and discard the galaxies in them.
As a result, we have a binary survey mask consisting of base pixels that are either empty, or fully covered.
Keeping track of the survey mask is now simple: we only need to keep track of which base pixels are included in the survey.

Discarding the galaxies at the survey edges means that we lose part of the information in the galaxy catalog.
In practice, the data loss is small and has little effect on the results.
The data loss can be reduced by increasing the base resolution, but a very high base resolution leads to inefficient data handling.

The data stored in memory consists of the number of galaxies in each cell of the 3D grid, and their co-added shear (if available).
In a typical situation, the grid is sparse and a majority of the cells are empty.
\fuga\ stores the data in a memory-efficient way, as follows.
For each base pixel under the mask, the code stores a two-dimensional data object, an array of arrays.
The first dimension corresponds to high-resolution pixels within the base pixel.
For each of those high-resolution pixels, the code stores an array of indices pointing to the redshift bins that contain a galaxy (or several).  
Other data objects of the same structure hold the number of galaxies and the co-added shear for the same cells.  
This data storage scheme ensures efficient use of memory, as empty cells are not kept in memory.

The hierarchical pixelization offers flexibility in handling different survey footprints. With the same code and with the same data structures
we can store a full-sky survey at modest pixel resolution, or a small survey with a very high resolution.

The discretization loses information on scales smaller than the pixel size.
The error can be reduced by increasing the resolution, at the price of increasing computational cost. There is thus a trade-off involved.
The results in this work have been obtained with resolutions in the range \tnside=512--2048 and \tzdelta=0.0005--0.005.

%%%%%%%%%%%%%%%%%%%%%
\subsection{Clustering correlation function}
\label{sec:clusteringCorrelation}

The standard way of estimating the correlation function of a galaxy sample is the Landy-Szalay estimator
\be
\xi(q) = \frac{dd(q)-2\cdot dr(q)}{rr(q)} +1
\ee
where $dd, dr, rr$ are normalized pair counts: galaxy-galaxy, galaxy-random, and random-random pairs. The estimator requires, by the side of the data catalog, a random catalog that represents an uncorrelated distribution with the same selection function and boundary effects as the actual data catalog. The parameter $q$ here represents a general coordinate that identifies a data bin. It may be taken to represent, for instance, an interval in radial distance $[r,r+\delta r[$, an interval in angular separation $[\theta,\theta+\Delta\theta[$, or, in our case, an element in the three-dimensional parameter space of $\{z_1,z_2,\theta\}$. The counts are normalized to the total number of pairs.

\fuga\ applies to correlation function estimation a method that mimics the Landy-Szalay estimator at the limit of an infinite random catalog. Consider one cell of our 3D grid at redshift $z$ and in angular direction $\Omega$. The cell combines a \healpix\ pixel with a small redshift interval. We denote the expected number of observed galaxies in the cell by $\alpha(z,\Omega)$. It can be written as
\be
\alpha(z,\Omega) = \bar n_{\rm tot}(z)S(z,\Omega)\Delta V(z),  \label{eq:alphaDef}
\ee
where $\bar n_{\rm tot}(z)$ is the mean galaxy density at redshift $z$, a function of $z$ to allow for time-evolution effects. 
The density of observable galaxies is $\bar  n=S(z,\Omega)\bar n_{\rm tot}$, where
 $S(z,\Omega)$ is the position-dependent selection function, and $\Delta V(z)$ is the volume of the grid cell. Since \healpix\ pixels have equal areas, the volume of the cell is proportional to the radial diameter of the cell. For a given $z$ bin it depends on the cosmology, and thus is not known a priori. Nor do we know the mean density or the selection function independently.
What we need, however, is not the individual factors, but the combination of Eq.~(\ref{eq:alphaDef}),  and this we can estimate from the data,
assuming that the selection function is either isotropic over the mask area, or a slowly varying function of $\Omega$. 
It is obtained from the average number of galaxies per pixel in the redshift shell and the area in question.
Obviously, the average must be taken over a sufficiently large area to smooth out local fluctuations.
By default, \fuga\ estimates $\alpha$ from the input galaxy catalog. Alternatively, if external information on the mean density is available,  this additional information can be injected in the form of a random catalog, which \fuga\ then converts into an estimate for $\alpha$.
 
From here on we assume that $\alpha(z,\Omega)$ is known. We denote by $\alpha(z_k,\Omega_{ij})$ its value on the $k$th redshift shell and in \healpix\ pixel $\Omega_{ij}$. We introduce here a double-index notation where the first index $i$ refers to a base pixel, and the second index $j$ labels high-resolution pixels inside it. The motivation for the double indexing scheme will become clear in due course. Similarly, we denote by $n(z_k,\Omega_{ij})$ the actual number of galaxies in the same cell.  Due to the chosen normalization 
\be
\sum_{ijk}\alpha(z_k,\Omega_{ij})=\sum_{ijk}n(z_k,\Omega_{ij})=N,
\ee
 where $N$ is the total number of galaxies in the sample.
The quantity $\alpha(\Omega_{ij},z_k)$ represents the number of galaxies in a 3D cell, if the galaxies
were uniformly distributed, and takes the role of the random catalog in the Landy--Szalay estimator.

The true RCF, as defined through Eq.~(\ref{eq:RCFdef1}), is a continuous function of its parameters.
When it is estimated from data, the parameter space must be divided into discrete bins (just as with the usual 2PCF).
We divide the range of interest in $\theta$ into $N_\theta$ bins, and label them as $\theta_m$, $m=1\ldots N_\theta$.
In redshift, we adopt the bin structure from the 3D space, that is, spacing at resolution \tzdelta.
In the following, a {\em cell} refers to an element of the spherical 3D grid, specified by coordinates $\{z,\Omega\}$, while a {\em bin} 
refers to an element in the parameter space of $\{z_1,z_2,\theta\}$.
  
We will need three data objects that correspond to the pair counts $dd,dr,rr$ of the Landy-Szalay estimator.
They are data objects of the same size and form as the RCF itself, that is, defined in the coordinate space $\{z_1,z_2,\theta\}$.
% $N_{\rm zbin}^2N_\theta$, where $N_\theta$ is the number of bins in $\theta$, and $N_{\rm zbin}$ is the number of redshift bins. The redshift grid coincides with that of the 3D data grid. 
 In the following, we consider in turn each of the objects $dd,dr,rr$. We start from $dd$, which is constructed as
 \begin{multline}
dd(z_k,z_{k'},\theta_m) =  \\ 
\frac{1}{N(N-1)} 
 \sum_{ii'\in S}  \sum_{jj'}  \Theta(\Omega_{ij}\!-\!\Omega_{i'j'}\in\theta_m) n(z_k,\Omega_{ij})n(z_{k'},\Omega_{i'j'})
 \label{eq:dd}
 \end{multline}
where $ii' \in S$ indicates that we include in the first sum only the base pixels inside the survey area, and
$\Theta(\Omega_{ij}-\Omega_{i'j'}\in\theta_m)$ has the role of assigning the pairs into the respective $\theta$ bins. 
Specifically, $\Theta(\Omega_{ij}-\Omega_{i'j'}\in\theta_m)=1$ if the angular distance between the centers of pixels $\Omega_{ij}$ and $\Omega_{i'j'}$ falls in bin $\theta_m$, and vanishes otherwise.
As implied by the prefactor $1/(N(N-1))$, we do not count pairs where both components are the same galaxy.
Thus, in the special case where $i=i',j=j',k=k'$, the term inside the sum should read $n(z_k,\Omega_{ij})(n(z_k,\Omega_{ij})-1)$, and this is assigned to the lowest $\theta$ bin. For clarity of expression, we do not write this separately in Eq.~(\ref{eq:dd}).

The ordering of the elements in Eq.~(\ref{eq:dd}) reflects the loop structure of the code.
Evaluating the expression of Eq.~(\ref{eq:dd}) algorithmically involves six nested loops, with the four outermost loops corresponding to the sums in Eq.~(\ref{eq:dd}), and the two innermost ones to the redshift bins. Let $N_{\rm base}$ be the number of pixels at base resolution, $N_{\rm sub}=$ the number of high-resolution pixels included in one base-resolution pixel, and $N_{\rm zbin}$ the number of redshift shells. For example, with \tnbase=32 and \tnside=1024 we have $N_{\rm base}=12\cdot32^2=12288$, and $N_{\rm sub}=(1024/32)^2=1024$. Naively, the computational complexity of the evaluation would be $N_{\rm base}^2\times N_{\rm sub}^2\times N_{\rm zbin}^2$. In reality, it is much less than that.
The first two loops($i,i'\in S$) scan through pairs of base pixels within the survey mask. Usually, one is interested in a limited angular range $\theta<\theta_{\rm max}$. Base pixel pairs far outside the range of interest can be discarded immediately (the \healpix\ package includes routines for this).  The actual number of base pixel pairs that require closer inspection is typically of the order of $N_{\rm pairs}\sim 10\,000-100\,000$ rather than $N_{\rm base}^2$. Thus, the two-level scheme works not only as a means for bookkeeping, but also as an efficient preselector in correlation function estimation.
The next two loops ($j,j'$) scan through high-resolution pixel pairs between the base pixel pair ($i,i'$). Most of the computation time is taken by the evaluation of the angular distance $\theta$ between the pixel pair.  This does not depend on redshift and only needs to be done once for a pixel pair. The two innermost loops scan over the redshift bins ($k,k'$), and accumulate the products $n(z_k,\Omega_{ij})n(z_{k'},\Omega_{i'j'})$ in the respective bins of the $dd$ grid. In a typical case, most redshift cells are empty. For instance, with \tnbase\ and with $N_{\rm zbin}=1000$ redshift bins we have over $10^{10}$ grid cells, while a typical galaxy catalog has a few million objects. It is thus inevitable that a vast majority of the cells are empty (and those that are not, mostly contain just one galaxy). Since only the non-empty grid cells are stored in memory, the last ($k,k'$) loops only need to scan through a couple of elements.
In total, the computational complexity of the evaluation of the $dd$ array is of the order of $N_{\rm pairs}\times N_{\rm sub}^2\times \text{(a few)}$.

We examine the evaluation of $rr$, which represents random-random pair counts. We do not have an explicit random catalog; its role is taken by the factor $\alpha$, which incorporates our knowledge of the survey geometry and of the selection function. Imagine that we had a very large random catalog with $MN$ objects, and $M\rightarrow\infty$. The number of random points in a cell approaches $M\alpha(z_k,\Omega_{ij})$, and the normalized $rr$ count is obtained as
%
%Just alternative formatting!
% \bea
%&& rr(z_k,z_{k'},\theta) = \nonumber  \\
%\lefteqn{\frac{1}{MN(MN-1)} \sum_{ijk}\sum_{i'j'k'}
% M\alpha(z_k,\Omega_{ij})M\alpha(z_{k'},\Omega_{i'j'}) T(\theta,\Omega_{ij},\Omega_{i'j'}) } \nonumber \\
%\lefteqn{ \rightarrow \frac{1}{N^2}
% \sum_{ijk}\sum_{i'j'k'}
% \alpha(z_k,\Omega_{ij})\alpha(z_{k'},\Omega_{i'j'}) T(\theta,\Omega_{ij},\Omega_{i'j'}) \,. } 
% \label{eq:rr}
% \eea
%
%
 \begin{multline}
rr(z_k,z_{k'},\theta_m) = \\
\frac{1}{MN(MN-1)} \sum_{ij}\sum_{i'j'}  \Theta(\Omega_{ij}\!-\!\Omega_{i'j'} \!\in\! \theta_m) 
 M\alpha(z_k,\Omega_{ij})M\alpha(z_{k'},\Omega_{i'j'})  \\
\rightarrow \frac{1}{N^2}
 \sum_{ij}\sum_{i'j'}  \Theta(\Omega_{ij}\!-\!\Omega_{i'j'}\in\theta_m) 
 \alpha(z_k,\Omega_{ij})\alpha(z_{k'},\Omega_{i'j'}) \,. 
 \label{eq:rr}
 \end{multline}
This is similar to the expression for $dd$, with the difference that instead of $n$ we have $\alpha$, which, unlike $n$, is not sparse. Thus, the evaluation with the $dd$ algorithm would be much more expensive.
This is where the assumption of an isotropic selection function steps in.
In the simplest case, the selection function can be assumed to be isotropic over the survey area, that is, it depends only on $z$. The isotropy of the selection function implies an isotropic survey depth: the survey shape is a cone, with equal depth in all directions within the survey area. We can then write $\alpha(z_k,\Omega_{ij})=\alpha(z_k)m(\Omega_{i})$, where $m(\Omega_{i})$ is a binary survey mask, with $m(\Omega_{i})=1$ when the base pixel is inside the survey area and $m(\Omega_{i})=0$ when the pixel is outside. Recall that our survey area has been trimmed to consist of full base pixels; therefore, the mask only depends on the index $i$.
The sum can now be rearranged to
 \be
rr(z_k,z_{k'},\theta_m) = 
\frac{1}{N^2}
\alpha(z_k)\alpha(z_{k'})
 \sum_{ii' \in S} \sum_{jj'}
\Theta(\Omega_{ij}-\Omega_{i'j'}\in\theta_m) \,.
 \label{eq:rr1}
 \ee
 The four sums simply count the number of pixel pairs with angular distance inside bin $\theta_m$.
 The computational complexity of this operation is $N_{\rm pairs}\times N_{\rm sub}^2$, and the result is a small array of length $N_\theta$. Scaling it by $\alpha(z_k)$ and $\alpha(z_{k'})$ and spreading it on the $rr$ grid is a very fast process.
 
In a somewhat more complicated scheme, we relax the assumption of isotropy over the survey area and require isotropy only within a base pixel. We can now write $\alpha(z_k,\Omega_{ij})=\alpha(z_k,\Omega_i)$. The expression for $rr$ becomes
\bem
rr(z_k,z_{k'},\theta_m) = \\
\frac{1}{N^2}
 \sum_{ii'\in S}\alpha(z_k,\Omega_i)\alpha(z_{k'},\Omega_{i'})
 \sum_{jj'}  \Theta(\Omega_{ij}-\Omega_{i'j'}\in \theta_m) \,.
 \label{eq:rr2}
 \end{multline}
 The procedure of counting pixel pairs, scaling by $\alpha$, and spreading on the target grid, is now repeated for each pair of base pixels $ii'$. This is computationally more expensive than the fully isotropic case, but still significantly less expensive than brute-force computation with an arbitrary selection function.

The third element in the correlation function estimation is $dr$, which represents cross-pair counts between the data catalog and the random catalog. 
Combining the characteristics of $dd$ and $rr$ estimation, we find
\bem
dr(z_k,z_{k'},\theta_m) = \\
\frac{1}{N^2}
\alpha(z_{k'}) \sum_{ii'\in S}\sum_j
 n(z_k,\Omega_{ij}) \sum_{j'}\Theta(\Omega_{ij}-\Omega_{i'j'}\in \theta_m) \,.
\end{multline}
for the isotropic case, and  
\bem
dr(z_k,z_{k'},\theta_m) = \\
\frac{1}{N^2}
\sum_{ii'\in S} \alpha(z_{k'},\Omega_{i'}) 
\sum_j n(z_k,\Omega_{ij}) \sum_{j'} \Theta(\Omega_{ij}-\Omega_{i'j'}\in \theta_m)  \,. 
 \label{eq:dr}
\end{multline}
for the pixelwise isotropic case.
The sum over $j'$ counts pixels within distance $\theta_m$ from pixel $\Omega_{ij}$. This is multiplied by the galaxy counts in the first pixel, accumulated in an array of size ($N_{\rm zbin},N_\theta$), and finally distributed on the $dr$ grid.

The cross-count $dr$ is not symmetric with respect to exchange $dr \leftrightarrow rd$ (a galaxy at low $z$, and a random point at high $z$, is not the same thing as a galaxy at high $z$, and a random point at low $z$). The final correlation function must therefore take into account both,
\bem
\hat\xi(z_k,z_{k'},\theta_m) =  \\
\frac{dd(z_k,z_{k'},\theta_m) -dr(z_k,z_{k'},\theta_m)-rd(z_k,z_{k'},\theta_m)}
{rr(z_k,z_{k'},\theta_m)} +1
\label{eq:clustering_combined}
\end{multline}
where
\be
rd(z_k,z_{k'},\theta_m) = dr(z_{k'},z_k,\theta_m) \,.  
\ee

When constructing the components of correlation function we exploit the symmetry with respect to the exchange $ijk\leftrightarrow i'j'k'$ to speed up the computation. The loops run through half of the index space, and we update two elements of the target array at one step. This involves having to keep track of the $dr$ and $rd$ arrays separately.

%%%
\subsection{Shear correlation function}
\label{sec:shearCorrelation}

Computing the correlation function for shear is rather straightforward compared with the computation of the clustering correlation. 
To start, the measured ellipticity of a galaxy is interpreted as a direct measure of the shear field $\gamma$ at the location of the galaxy.
The shear of a 3D cell, $\gamma_{ijk}$, is taken to be the average of all galaxy ellipticities in that cell,
\be
\gamma_{ijk} \equiv \gamma(z_k,\Omega_{ij}) =
\begin{cases} 
 n(z_k,\Omega_{ij})^{-1} \sum_{m} \epsilon_m,  & n(z_k,\Omega_{ij})>0\\
 0, & n(z_k,\Omega_{ij})=0
 \end{cases}
\ee
where the sum is over all galaxies in the cell defined by pixel $\Omega_{ij}$ and redshift $z_k$, and $n(z_k,\Omega_{ij})$ is their total number.
\fuga\ evaluates the shear correlation as a weighted sum over pairs of 3D cells,
\bem
\hat\xi_\pm(z_k,z_{k'},\theta_m) = \label{eq:shearsum} \\
\frac{\sum_{ii'\in S} \!\sum_{jj'}\! \Theta(\Omega_{ij}\!-\!\Omega_{i'j'} \!\in\! \theta_m)  w_{ijk}w_{i'j'k'}
(\gamma_{ijk}^{\rm t} \gamma_{i'j'k'}^{\rm t} \!\pm \gamma_{ijk}^\times\gamma_{i'j'k'}^\times) }
{\sum_{ii'\in S}\sum_{jj'} \Theta(\Omega_{ij}\!-\!\Omega_{i'j'} \!\in\! \theta_m)  w_{ijk}w_{i'j'k'} }
\end{multline}
where $w_{ijk}$ is a weight associated with cell $ijk$. 
\fuga\ implements two weighting schemes: galaxy weighting $w_{ijk}=n(z_k,\Omega_{ij})$ (all galaxies carry the same weight), and pixel weighting where $w_{ijk}=1$ for all cells that contain at least one galaxy, and $w_{ijk}=0$ for the empty ones.  Galaxy weighting is equivalent to estimating the shear correlation as a sum over galaxy pairs, as is the usual procedure \citep{Schneider:2002jd,Kilbinger:2014cea}. At very high resolution, when there is just one galaxy in a cell (and the majority of cells are empty), the two weighting schemes become equivalent.
Note that we cannot apply fully uniform weighting, since we cannot define ellipticity in a meaningful way for empty cells.

Above, $\epsilon^{\rm t}$ and $\epsilon^\times$ are the tangential and cross-components of the measured ellipticity (see e.g. \citealt{Kilbinger:2014cea}).
The rotation into tangential and cross components is done with respect to the great circle connecting the pixels in question, and must be recomputed for every pixel pair. The estimation of the correlation function involves again six nested loops. The outer four ($i,i',j,'j'$) scan the sky pixels, the two inner ones the redshift shells. For a given pair of pixels, we rotate the shear of all the objects within those pixels to their tangential and cross components, as this operation only depends on the pixel position, not on redshift. Looping over redshifts and updating the sum of Eq.~(\ref{eq:shearsum}) is then a fast process.

%%%%%%
\subsection{Galaxy-shear correlation}

To fully exploit the information in a galaxy catalog, we also compute the cross-correlation between the galaxy density and the shear field.  The density fluctuation  $\delta$ is defined as the relative deviation from uniform density,
\be
\delta(z_k,\Omega_{ij}) = \frac{n(z_k,\Omega_{ij})-\alpha(z_k,\Omega_{ij})}{\alpha(z_k,\Omega_{ij})}  . \label{eq:deltaDefObs}
\ee
The cross-correlation between the density fluctuation and the tangential shear is constructed as
\bem
\hat\xi_{\rm ct}(z_k,z_{k'},\theta_m) = \\
\frac{\sum_{ii'\in S}\sum_{jj'} \Theta(\Omega_{ij} \!-\! \Omega_{i'j'} \!\in\! \theta_m)  w_{ijk} \alpha(z_k',\Omega_{i'j'})
\gamma_{ijk}^{\rm t} \delta(z_{k'},\Omega_{i'j'}) }
{\sum_{ii'\in S}\sum_{jj'} \Theta(\Omega_{ij} \!-\!\Omega_{i'j'} \!\in\! \theta_m) w_{ijk}\alpha(z_{k'},\Omega_{i'j'}) } 
\label{eq:crossSum}
\end{multline}
This is a weighted sum of products $\epsilon^{\rm t}\delta$,
where the weights are $w_{ijk}$ for the shear component, and $\alpha$ for the density component.
With Eq. (\ref{eq:deltaDefObs}), this breaks into a combination of three sums,
\be
\hat\xi_{\rm ct}(z_k,z_{k'},\theta_m) =
\frac{A(z_k,z_{k'},\theta_m) -B(z_k,z_{k'},\theta_m)}{C(z_k,z_{k'},\theta_m)}
\ee
where
\bea
A(z_k,z_{k'},\theta_m) &=&  \sum_{ii'\in S}\sum_{jj'} \Theta(\Omega_{ij}\!-\!\Omega_{i'j'} \!\in\! \theta_m) 
w_{ijk}\gamma_{ijk}^{\rm t} n(z_{k'},\Omega_{i'})\nn
B(z_k,z_{k'},\theta_m) &=& \sum_{ii'\in S}\alpha(z_{k'},\Omega_{i'})\sum_{jj'} \Theta(\Omega_{ij}\!-\!\Omega_{i'j'} \!\in\! \theta_m) 
w_{ijk}\gamma_{ijk}^{\rm t}  \nn
C(z_k,z_{k'},\theta_m) &=& \sum_{ii'\in S}\alpha(z_{k'},\Omega_{i'})\sum_{jj'} \Theta(\Omega_{ij}\!-\!\Omega_{i'j'} \!\in\! \theta_m) 
w_{ijk} .
\eea
Here we used the assumption that the reference density $\alpha$ is uniform within a base pixel, and moved it out from the sum over $j$.
Should $\alpha$ be isotropic and a function of $z$ only, it can further be moved out from the sum over $i$, in the same way as in the computation of $dr$.

The procedure of constructing $A(z_k,z_{k'},\theta_m)$ is equivalent to the procedure of constructing $dd(z_k,z_{k'},\theta_m)$ in Section \ref{sec:clusteringCorrelation}, just replacing $n(z_k,z_{k'},\theta_m)$ with $w_{ijk}\epsilon_{ijk}$. The procedures for $B(z_k,z_{k'},\theta_m)$ and $C(z_k,z_{k'},\theta_m)$ are equivalent to that for $dr$.
We thus already have the machinery in place to construct the cross-correlation between shear and the clustering signal.

In the same way, we construct estimates for the cross-correlation components $\xi_{\rm c\times},\xi_{\rm t\times},\xi_{\times\rm t}$.
While parity symmetry suggests that these spectra will vanish, their estimates constructed from data will differ from zero randomly, due to cosmic variance and measurement errors. The ``zero'' spectra thus contain valuable information on the error level in the other spectra.
\fuga\ by default computes all the spectral components, as leaving the zero components out does not significantly affect the computation time.

%%%%%%%
\subsection{Rebinning }

\fuga\ stores the RCF on disk as a FITS file.  
The parameters that control the size and shape of the table are:  redshift limits $z_{\rm min},z_{\rm max}$, maximum angular separation $\theta_{\rm max}$ and the number of bins $N_\theta$, and maximum redshift separation.
The redshift resolution \zdelta (set by \tzdelta) is inherited from the grid resolution. 
For each combination $\{z_k,z_{k'},\theta_m\}$, the code stores the components of the correlation function, and the corresponding weighting factors.  
Storing the weights along with the correlation function is crucial for the re-binning of the RCF.
In the case of clustering, the weight is $rr(z_1,z_2,\theta)$, the denominator in Eq.~(\ref{eq:clustering_combined}).
When we multiply the correlation function by the denominator, we restore the additive nominator, and we can re-bin it to an alternative bin structure.
For example, the RCF can be rebinned to form the angular correlation function $\xi(\theta)$ for an arbitrary redshift shell $[z_1,z_2]$ 
as
\be
  \xi(\theta) = \frac{\sum_{z_k,z_{k'}\in[z_1,z_2]} \xi(z_k,z_{k'},\theta) rr(z_k,z_{k'},\theta)} {\sum_{z_k,z_{k'}\in[z_1,z_2]} rr(z_k,z_{k'},\theta)} \,.
\ee
Weighting by $rr$ ensures that the resulting correlation function is the same as if the $dd,dr,rr$ pairs were initially counted in the new binning.
Similar re-binning schemes can be applied to shear and cross-correlation.
The weighting factor for shear is the denominator of Eq.~(\ref{eq:shearsum}), and that of cross-correlation the denominator of Eq.~(\ref{eq:crossSum}). All of these are stored in the RCF file along with the correlation function. 

%%%%%%%
\subsection{From redshift-space correlation to real space}

To transform the redshift-space correlation function $\xi(z_1,z_2,,\theta)$ into the conventional real-space correlation function, we employ a dedicated \realspace\ module that ingests outputs from \fuga, and converts the redshifts and angles into physical distances. Input and control parameters include one or several RCF files, redshift-shell definitions ($z_{\min},\,z_{\max}$),  cosmological parameters ($H_0$ and $\Omega_m$), and real-space binning settings ($r_{\max},\,N_r,\,N_\mu$). Here, $N_r$ and $N_\mu$ give the number of bins in separation distance $r$ and in $\mu$.
We define LOS as the radial vector passing through the midpoint of the separation vector.
Users can enable production of 1D correlation functions $\xi_0(r)$, 2D correlations $\xi(r,\mu)$, or Legendre multipoles $\xi_\ell(r)$ ($\ell=0,2,4$).

The redshift coordinates $(z_1, z_2)$ are converted to comoving distances via either a Hubble-law approximation or a pre-tabulated flat $\Lambda$CDM interpolation. Data are filtered by redshift shell, and projection routines convert the $\{ z_1,z_2,\theta\}$ coordinates into real-space separations $r$ and orientation angles $\mu$, re-bin the elements of the input RCF into $\xi(r,\mu)$ of $\xi(r)$ with $rr$ weighting, and, when requested, project onto Legendre polynomials to extract multipoles.  Because of the discretization, where galaxies are assigned to the centers of their respective grid cells, it easily happens that one or several bins of the $\{r,\mu\}$ grid are empty, and the $\xi(r,\mu)$ correlation function remains undefined. For this reason we do not evaluate the multipoles $\xi_\ell(r)$ by direct integration, but perfom a linear fitting of $\ell=0,2,4$ Legendre polynomials $L_\ell(\mu)$ to $\xi(r,\mu)$ for each $r$, using again the $rr$ count as weight. The fit coefficients determine the multipoles $\xi_\ell(r)$. 

The \realspace\ module can optionally be run in statistical mode, where the code ingests multiple RCF files. An online \texttt{CovAccumulator} class implements Welford's algorithm to aggregate means and covariance matrices across all realizations, writing FITS HDUs: one binary table for the mean correlation and a companion image HDU for its covariance.  This is helpful when analyzing simulations that provide a large number of realizations. 
\realspace\ also allows scanning over a range of cosmological parameters $H_0,\Omega_m$ in one run.

\begin{figure}
\includegraphics[width=1\linewidth]{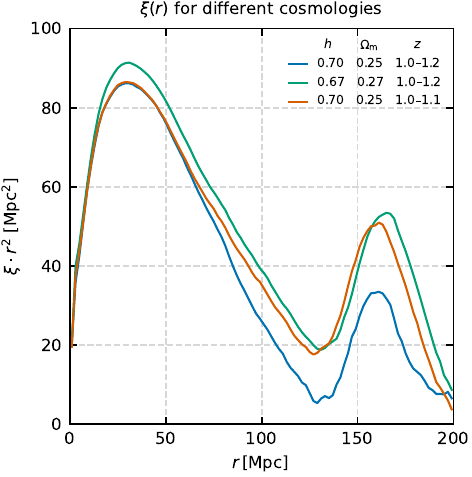}
\caption{Real-space 2PCF $\xi(r)$ from MICE simulation, for different assumed cosmologies and redshift ranges.
The correlation functions have been constructed from the same input RCF.
\label{fig:realcorr}}
\end{figure}

The use of \realspace\ is demonstrated in Fig.~\ref{fig:realcorr}.
Here we have computed the correlation function monopole for two different assumed cosmologies, and for different redshift ranges, from the same input RCF file.
Once the RCF file is available, constructing different variants of the real-space correlation function takes a fraction of a second with the \realspace\ tool,
and can easily be run on a laptop.

%%%%%%%
\section{Harmonic analysis}
\label{sec:harmonic}

The spherical 3D grid forms a natural base for harmonic analysis.
%The harmonic-space counterpart of the RCF is the redshift-space power spectrum.
\fuga\ provides the option of computing the spherical harmonic transform (SHT) of the density field and the shear field, and constructing angular power spectra. 
Doing this by redshift bins leads to the redshift-space power spectrum (RP) $P_\ell(z_1,z_2)$, which is the harmonic counterpart of the RCF.

\fuga\ computes the SHT using the exact coordinates (RA, DEC) of the input galaxies. The coordinate space is discretized only in the redshift dimension. The power spectrum is thus independent of \healpix\ resolution.  Power-spectrum analysis using discrete coordinates has been discussed thoroughly by \cite{Euclid:2024xqh} (hereafter TS25) in a recent work.
The difference between our work and that of TS25 arises from the different handling of the reference density. We take the reference galaxy density to be a known function, which can be evaluated at each galaxy position, while in TS25 it enters through a random catalog, which is treated with the same discrete SHT as the actual catalog. In the latter approach, it is not possible to directly transform the relative density fluctuation $\delta$, since there is no such thing as a catalog divided by another catalog. As a consequence, TS25 take as the starting point the SHT of the density field itself, while we analyze the relative fluctuation $\delta$.

\subsection{SHT with discrete coordinates}

We now proceed to describe the discrete SHT, as implemented in \fuga.
Consider first the SHT of the density contrast in one redshift bin $z_k$.
We want to evaluate the integral
\be
a_{C,\ell m}(z_k) = \int_S \delta_{2D}(\vartheta,\varphi;z_k)
Y^\ast_{\ell m} (\vartheta,\varphi) d\Omega \,,
\ee
where $\int_S$ indicates that the integral is carried out over the survey area, and $\delta_{2D}$ is a 2-dimensional density fluctuation on the sphere, defined as
\be
\delta_{2D}(\vartheta,\varphi;z_k) = \frac{\sigma(\vartheta,\varphi;z_k)}{\bar\sigma(\vartheta,\varphi;z_k)} -1.
\ee
Here  $\sigma(\vartheta,\varphi;z_k)$ is the angular number density, number of objects per unit solid angle in direction $(\vartheta,\phi)$ in redshift bin $z_k$,
and $\bar\sigma(\vartheta,\varphi;z_k)$ is the corresponding background quantity.
%It represents an uncorrelated density field with the same selection function and survey shape as the actual density field.
As before, we allow for it to depend on location, to accommodate general selection effects.
Quantity $\alpha$ defined in the Sect. \ref{sec:clusteringCorrelation} can be understood as an integral of $\bar\sigma$ over the area of one pixel.
%\fuga\ offers the option of estimating an isotropic $\bar\sigma(z_k)$ from the input catalog, or from an external random catalog.
From here on, we assume that the background density is known.

Since galaxies are discrete objects, the angular density $\sigma(\vartheta,\varphi;z_k)$ is formally a combination of delta peaks at the locations of the galaxies.
Taking an integral over delta peaks converts the integral into a sum over galaxy positions,
\be
a_{C,\ell m}(z_k) =  \sum_i w_i Y^\ast_{\ell m} (\vartheta_i,\varphi_i) 
- \int_S Y^\ast_{\ell m} (\vartheta,\varphi) d\Omega \,.
\label{eq:clusteringSHT}
\ee
The weights $w_i$ are taken from the values of $\bar\sigma$ at the locations of the galaxies,
\be
  w_i = \frac{1}{\bar\sigma(\vartheta_i,\varphi_i;z_k)}.
\ee
For an isotropic selection function $w_i=4\pi f_{\rm sky}/N$, where $N$ is the total number of objects in the survey area and $f_{\rm sky}$ is the sky fraction.

Computing the sum of Eq.~(\ref{eq:clusteringSHT}) by brute force is an expensive operation, but a very efficient approximation is possible by making use of a combination of traditional SHT and non-uniform Fast Fourier Transforms (NUFFTs), as described by \cite{Reinecke:2023gtp}.
To recapitulate the essential ingredients of this operation very briefly, an SHT from spherical harmonic coefficients to an arbitrary point set on the sphere can be broken down into the following steps:
\begin{enumerate}
    \item conventional SHT from coefficients onto an equiangular grid of sufficient resolution for the given band limit
    \item resolution increase of the resulting map by FFTs + zero-padding in both angular directions
    \item ``Fourier interpolation'' from grid points to non-equidistant points.
\end{enumerate}
Naturally, the adjoint of such an operation can be carried out by reversing the order of the steps, but it is important to realize that a genuine \emph{inverse} operation is (in contrast to some regular grids) generally not possible.

A spherical harmonic transform involving arbitrarily distributed points on the sphere can be implemented to run at speeds comparable to one working on a regular grid like HEALPix, Gauss-Legendre, etc., with the important caveat that the transform results will not be intrinsically accurate to (nearly) machine precision like standard SHTs, but are approximations with an error that can be tuned by the user. Accuracies close to machine precision can be reached, but naturally require higher computational resources (in terms of both CPU time and memory) than less accurate transforms.

The sum in eq.~(\ref {eq:clusteringSHT}) perfectly matches the adjoint general spherical harmonic transform described in detail in section 3 of
\cite{Reinecke:2023gtp}.
Its run-time requirement scales according to
\begin{equation}
t_\text{SHT} = \mathcal{O}(\lmax^3) + \mathcal{O}(N)\text{.}
\end{equation}
The first term is the cost of the underlying standard SHT onto an equiangular grid capable of representing moments up to \lmax\ without loss, and comparable to the cost of an SHT to or from a HEALPix grid with $\nside = \lmax/2$. The second term scales with the number of non-uniformly spaced points (in this case individual galaxies) and can be roughly estimated as $10^{-7}\text{s}$ per point on a single CPU core at an accuracy close to machine precision.

It is interesting to note that, depending on the choice of band limit and number of points, either of the two terms in the above equation may dominate over the other. Assuming a subdivision of the data into many redshift shells and a high band limit, the $\mathcal{O}(\lmax^3)$ term will account for almost the entire runtime, while for a single SHT including all galaxies at intermediate \lmax, the situation will be reversed.

The second term in (\ref {eq:clusteringSHT}) represents an SHT of the survey mask and is computed as a conventional map-based SHT, with a resolution that depends on the chosen maximum $\ell$. The \healpix\ resolution parameter $N_{\rm side}$ is taken to be the smallest power of two such that $N_{\rm side}$ exceeds $\ell_{\rm max}$.
For instance, with $\lmax=4000$ the transform is computed from a $N_{\rm side}$=4096 map.

The redshift-space pseudo-spectrum for clustering is constructed as
\be
\hat C^{CC}_\ell(z_k,z_{k'}) = \frac{1}{2\ell+1}\sum_m | a^\ast_{C,\ell m}(z_k) a_{C,\ell m}(z_{k'}) | \, .
\label{eq:pseudoCCC}
\ee
Because it is computed from a field with incomplete sky coverage, the pseudo-spectrum underestimates the true underlying power spectrum. As a first approximation,
\be
\hat C^{CC}_\ell(z_k,z_{k'}) \approx f_{\rm sky}C^{CC}_\ell(z_k,z_{k'}) ,
\label{eq:skyScaling}
\ee
where $f_{\rm sky}$ is the sky fraction and $C^{CC}_\ell(z_k)$ is the true spectrum (e.g. \citealt{Hivon:2001jp}). A more accurate sky correction, based on a coupling kernel, is discussed in Section \ref{sec:skymask}.

The SHT of the shear field involves spin-weighted spherical harmonics,
\be
a_{\pm2,\ell m}(z_k) = \frac{4\pi}{N} \sum_{i} [\gamma_{i1}\pm i\gamma_{i2} ]
\,{}_{\pm2}Y_{\ell m} (\vartheta_i,\varphi_i) \,,
\ee
where again the sum is over objects in redshift bin $z_k$. The normalization is chosen so that for objects uniformly distributed over the full sky, this matches the continuous SHT of the underlying shear field.

The SHT coefficients can be combined into $E$ and $B$ components
\be
a_{E,\ell m} = -\half(a_{2,\ell m}+a_{-2,\ell m}), \quad
a_{B,\ell m} = -\tfrac{1}{2i}(a_{2,\ell m}-a_{-2,\ell m}), 
\ee
From these, one can construct angular power spectra for the shear, and a cross-spectrum between shear and the clustering signal,
\bea
\hat C^{EE}_\ell(z_k,z_{k'}) &=& \frac{1}{2\ell+1}\sum_m | a^\ast_{E,\ell m}(z_k)  a_{E,\ell m}(z_{k'}) |  \nn
\hat C^{BB}_\ell(z_k,z_{k'}) &=& \frac{1}{2\ell+1}\sum_m | a^\ast_{B,\ell m}(z_k)  a_{B,\ell m}(z_{k'}) |  \nn
\hat C^{CE}_\ell(z_k,z_{k'}) &=& \frac{1}{2\ell+1}\sum_m | a^\ast_{C,\ell m}(z_k)  a_{E,\ell m}(z_{k'}) | \,.
\eea
These are the ``non-zero'' spectral components.
The ensemble averages of the remaining combinations, ($EB$, $CB$), are expected to vanish on grounds of parity symmetry, but due to measurement errors and the random nature of the shear and density fields the measured spectra will differ from zero, and they contain valuable information on the error level in the other spectra. 
\fuga\ computes and stores all the spectral components.

%%%%%%%
\subsection{Sky mask}
\label{sec:skymask}

The pseudo-spectrum estimated on a partial sky gives a biased estimate of the true spectrum of the field.
%As a first approximation, the effect of the sky mask is to scale the spectrum down by the sky fraction, as by Eq.~(\ref{eq:skyScaling}). 
The survey footprint also introduces couplings between multipoles. At the ensemble-average level, the effect of the sky mask is encompassed in a coupling kernel \citep{Hivon:2001jp,Camphuis:2022utm,Brown:2004jn,Challinor:2004pr}, which connects the true spectrum $C_\ell$ to the ensemble average of the pseudo-spectrum $\langle\hat C_\ell\rangle$. The coupling mixes $EE$ with $BB$, but, for instance, $EE+BB$ does not mix with $EE-BB$. We can arrange the non-zero spectra into independent combinations as follows:
\bea
\langle\hat C^{CC}_\ell \rangle &=& \sum_{\ell'} K^0_{\ell\ell'} C^{CC}_{\ell'} \nn
\langle\hat C^{CE}_\ell \rangle &=& \sum_{\ell'} K^\times_{\ell\ell'} C^{CE}_{\ell'} \nn
%\langle\hat C^{CB}_\ell \rangle &=& \sum_{\ell'} K^\times_{\ell\ell'} C^{CB}_{\ell'} \nn
\langle\hat C^{EE}_\ell +\hat C^{BB}_\ell \rangle &=& \sum_{\ell'} K^+_{\ell\ell'} (C^{EE}_{\ell'}+C^{BB}_{\ell'} ) \nn
\langle\hat C^{EE}_\ell -\hat C^{BB}_\ell \rangle &=& \sum_{\ell'} K^-_{\ell\ell'} (C^{EE}_{\ell'}-C^{BB}_{\ell'} ) \,.
%\langle\hat C^{EB}_\ell +\hat C^{BE}_\ell \rangle &=& \sum_{\ell'} K^-_{\ell\ell'} (C^{EB}_{\ell'}+C^{BE}_{\ell'} ) \nn
%\langle\hat C^{EB}_\ell -\hat C^{BE}_\ell \rangle &=& \sum_{\ell'} K^+_{\ell\ell'} (C^{EB}_{\ell'}-C^{BE}_{\ell'} ) \, .
\label{eq:Krelation}
\eea
These hold for each pair $z_k,z_{k'}$. We have dropped the arguments $z_k,z_{k'}$ for clarity.

The four types of kernel matrix are
\bea
K^0_{\ell\ell'} &=& \frac{ (2\ell'+1)}{4\pi}\sum_{\ell''} W^{\rm cc}_{\ell''}(2\ell''+1)
   \begin{pmatrix} \ell & \ell' & \ell'' \\ 0 & 0 & 0 \end{pmatrix}^2 \nn
K^\times_{\ell\ell'} &=& \frac{ (2\ell'+1)}{4\pi}\sum_{\ell''} W^{\rm cs}_{\ell''} (2\ell''+1)
   \begin{pmatrix} \ell & \ell' & \ell'' \\ 2 & -2 & 0 \end{pmatrix} 
   \begin{pmatrix} \ell & \ell' & \ell'' \\ 0 & 0 & 0 \end{pmatrix} \nn
K^+_{\ell\ell'} &=& \frac{ (2\ell'+1)}{4\pi}\sum_{\ell''} W^{\rm ss}_{\ell''} (2\ell''+1)
   \begin{pmatrix} \ell & \ell' & \ell'' \\ 2 & -2 & 0 \end{pmatrix}^2 \\
 K^-_{\ell\ell'} &=& \frac{ (2\ell'+1)}{4\pi}\sum_{\ell''} W^{\rm ss}_{\ell''} (2\ell''+1)
   \begin{pmatrix} \ell & \ell' & \ell'' \\ 2 & -2 & 0 \end{pmatrix}^2 (-1)^{\ell+\ell'+\ell''} \,, \nonumber
\eea
where $W^{ab}_\ell$, $a,b=$c,s, are the spectra of the sky mask,
\be
W^{ab}_\ell(z_k,z_{k'}) = \frac{1}{2\ell+1}\sum_m 
\omega^{a\ast}_{\ell m}(z_k)\omega^b_{\ell m}(z_{k'})
\ee
and $\omega^{\rm c}_{\ell m}(z_{k})$, $\omega^{\rm s}_{\ell m}(z_{k})$ are the SHT of the sky mask in $k$th shell, for clustering and shear, respectively.
The derivation for these formulas, including the zero components, is given in Appendix \ref{appendix:kernel}.
Our results for shear agree with those of \cite{Challinor:2004pr}, but there appears to be a sign difference with respect to \cite{Brown:2004jn}.

To construct the kernels, we need the SHT of the sky mask. 
There is a fundamental difference in how the sky mask is defined for the density fluctuation, or for the shear.
From shear, we only have information at the exact locations of the galaxies, and thus the mask consists of delta peaks at the locations of the same galaxies. The SHT of the shear mask becomes a discrete sum,
\be
\omega^{\rm s}_{\ell m}(z_k) = \frac{4\pi}{N}\sum_{i\in z_k} Y^\ast_{\ell m}(\vartheta_i,\varphi_i) \,,
\ee
computed with the same non-uniform SHT technique as the discrete part of the SHT of the density contrast.
The density field is observed over the full survey area; not observing a galaxy in a given location is a measurement of negative density contrast at that location.
The mask for the clustering signal is a continuous function, unity in the observed area, and zero outside it. The mask SHT is equivalent to the integral component of Eq.~(\ref{eq:clusteringSHT}),
\be
\omega^{\rm c}_{\ell m}=\int_S Y^\ast_{\ell m}(\vartheta,\varphi)d\Omega \,.
\ee

In a cone-like survey geometry, the clustering mask is the same for all redshifts.
Then it suffices to construct a single coupling kernel that applies to all $z_k,z_{k'}$ pairs. 
The situation is very different for the shear mask, which depends on the distribution of the galaxies, and is thus necessarily different for every $z_k,z_{k'}$ pair. With $N_{\rm zbin}$ redshift bins, we must construct $N_{\rm zbin}(N_{\rm zbin}+1)/2$ coupling kernels. This is a computationally heavy process, and we have paid special attention to the efficient construction of multiple kernels.

The computation time for the coupling kernels is dominated by the evaluation of the Wigner $3j$ symbols, but \fuga\ takes several steps to improve run-time in comparison to other existing implementations:
\begin{itemize}
    \item instead of processing a mask spectrum at a time, several coupling kernels are computed from several mask spectra simultaneously, avoiding the need to recompute the full set of Wigner $3j$ symbols over and over (pre-computing and storing them is unfortunately infeasible at the required band limits due to memory constraints).
    \item in a similar vein, the different kinds of kernels (0, $\times$, $-$, $+$) are also computed together, further improving the re-use of Wigner symbols
    \item the commonly used code by \cite{schulten:1975} was re-implemented with support for CPU SIMD instructions, allowing simultaneous computations of several coefficient vectors (2, 4, or 8, depending on the vector register width of the target CPU) at almost the same speed as a single one.
\end{itemize}
In combination, these improvements accelerate the generation of mode-coupling kernels by more than an order of magnitude in many typical circumstances, and therefore could also be of interest for similar numerical codes in the field.

If the survey covers a large enough fraction of the sky, it is possible to invert the kernel matrix and obtain an unbiased estimate of the true spectrum.
When the galaxy sample covers a small fraction of the sky, the kernel matrix may become singular. In that case, inverting the relations of Eq.~(\ref{eq:Krelation}) directly is not possible. The cure for this is to re-bin the power spectra and the kernel to wider bins. \fuga\ implements this as follows. Let matrix $B$ denote the simple binning operation, where $M_{\rm b}$ adjacent multipoles are co-added to produce an array of $N_{\ell\rm bin}=(\lmax+1)/M_{\rm b}$ elements, $B_{L\ell}=1$ when $\ell\in L$, and $L=1\cdots N_{\ell\rm bin}$ labels the wide bins. Assuming that the true spectrum $C$ is smooth, we can approximate it with a coarser version of the same spectrum with $N_{\ell\rm bin}$ elements,  denoted $\widetilde C_L$, and write $C_\ell=\sum_L B_{L\ell}\widetilde C_L$. 
Here $C,\widetilde C$ can be taken to represent any of the components in Eq.~(\ref{eq:Krelation}). 
The role of $B_{L\ell}$ here is to pick from the coarse spectrum the element $L$ to which $\ell$ belongs. 
We can now write
$\langle\hat C_\ell \rangle = \sum_{\ell'}K_{\ell\ell'} \sum_L B_{L\ell} \widetilde C_L$, and
multiplying both sides by $(2\ell+1)$ and binning, we arrive at
\be
\sum_{\ell\in L} (2\ell+1) \langle \hat C_\ell \rangle = \sum_{\ell\in L}(2\ell+1)\sum_{\ell'}K_{\ell\ell'} \sum_{L'}B_{L\ell} \widetilde C_L \,.
\ee
In matrix formalism, this reads
\be
B(2\ell+1) \langle \hat C^{CC} \rangle =B(2\ell+1)K^0 B^T \widetilde C^{CC} \,,
\ee
where $(2\ell+1)$ is to be read as a diagonal matrix, with factor $(2\ell+1)$ on the diagonal.
Taking $\hat C$ for $\langle\hat C\rangle$ we can now solve for the true spectrum as
\be
\widetilde C = [B(2\ell+1)K B^T]^{-1 } B(2\ell+1)  \hat C.
\ee
Matrix $B(2\ell+1)K B^T$ is the binned kernel matrix, of size $(N_{\ell \rm bin},N_{\ell\rm bin})$. To construct the binned kernel, we scale the initial kernel by $(2\ell+1)$ from the left, and bin the rows and columns, in the same way we bin the spectrum. Note that, unlike the initial kernel, the binned kernel is symmetric. On the right-hand side, we have the operation of scaling the pseudospectrum by $(2\ell+1)$ and co-adding. Scaling by $(2\ell+1)$ just cancels the division by $(2\ell+1)$ in Eq.~(\ref{eq:pseudoCCC}), thus we end up taking the direct sum of products of $a_{\ell m}$ pairs. This yields a natural and intuitive binning procedure.

The binning width required to make the kernel non-singular depends on the survey geometry. In the test cases of this work, binning over five adjacent multipoles has been sufficient to make the binned kernel matrix safely invertible. The bin width $M_{\rm b}$ is set by the user-defined parameter \texttt{rebin\_width}.  Since the spectrum length $(\lmax+1)$ is more often than not divisible by any natural binning width (4, 5, 10, 20), but $\ell_{\rm max}$ often is, we treat the monopole as a single bin, and start the actual rebinning from $\ell=1$.
The binned kernel is then inverted to yield an unbiased estimate of $\tilde C$. If needed, this can be expanded to full length as $B^T\tilde C$.

%%%%%%%
\subsection{From power spectrum to correlation function}

The power spectrum $P_\ell(z_k,z_{k'})$ can be converted into a correlation function. 
This provides an alternative way of constructing the RCF.
The relations connecting the RP to RCF involve reduced Wigner functions $d^\ell_{ss'}(\theta)$ for $s,s'=0,\pm2$. The explicit relations were given in Sect. \ref{sec:powspectheory}.

The process of computing the RCF from RP consists of the following steps:
\begin{itemize}
\item{Compute the pseudo-spectrum $\hat C_\ell$ using exact galaxy coordinates}
\item{from auto-components, subtract an estimate for the shot noise/white noise contribution}
%\item{Rebin the spectra to wider bins.}
\item{construct the coupling kernel $K$ and apply sky correction}
%\item{Expand back to original length.}
\item{convert to correlation function.}
\end{itemize}
This procedure is implemented in a post-processing tool of the \fuga\ code package.
The tool reads in an RP and the accompanying mask spectra from a file, and constructs the RCF.

%%%%%%%%%%%%%%%%%%%%%%%
\section{Validation}

\subsection{Test cases}

In the following, we describe the validation tests we performed on \fuga. As test data, we used the MICE Grand Challenge simulation, version 1 \citep{Fosalba:2013wxa,Crocce:2013vda,Fosalba:2013mra}.
MICE is a N-body simulation with 70 billion dark matter particles. From the available data products, we selected the light-cone simulation that spans an octant (1/8) of the sky, up to redshift $z=1.4$. The cosmological parameters used to create the simulation were $\Omega_m=0.25, \Omega_\Lambda=0.75, \Omega_b=0.044, n_s=0.95, \sigma_8=0.80, h=0.7$. 
Simulated galaxy catalogs are available through the \texttt{CosmoHub}\footnote{https://cosmohub.pic.es}  portal. We extracted the galaxy catalog with the following column information: \texttt{ra, dec, z, gamma1, gamma2}.
We used two versions of the galaxy catalog for our tests: the full-size catalog contains 205 million objects. We refer to this as ``full MICE''. For quick tests, we used a subset of 12.5 million objects (1/16 of the full catalog). To this we refer to as ``small MICE''.

For most of the tests, we used a subcatalog in the redshift range $z=1.0-1.2$. The full (small) MICE catalog contains 46.1 (2.89) million objects in this range. The edge-trimming procedure described in Section \ref{sec:resolution} reduced this further to 44.9  (2.81) million.
%46\,122\,887 (2\,885\,360) 44\,942\,934 (2\,811\,183).

We varied two parameters: the angular resolution \nside, set by the user parameter \tnside, and the redshift resolution \zdelta, set by parameter \tzdelta. The base pixel size was kept constant at \tnbase=128, to ensure that the sky coverage was the same in all runs. 

Unless stated otherwise, the tests were run on a single node of the {Puhti} supercomputer of the CSC -- IT Center for Science (Finland). {Puhti} is a supercluster with 682 CPU nodes. Each node is equipped with two Intel Xeon Gold 6230 processors with 20 cores each. 
Some smaller tests were run on a MacBook laptop with 8 cores.

\subsection{Comparison with full pair counting}

As the first validation test, we compared the two-point correlation function computed by \fuga\ against that computed by the public \corrfunc\footnote{https://corrfunc.readthedocs.io/en/master/}  code \citep{2020MNRAS.491.3022S}.
As input we had the full MICE catalog in the redshift range $z\in[1.0,1.2]$.
We did three runs with \fuga with increasing grid resolution, to first compute the RCF.  We used pixel resolution \nside=512, 1024, 2048,  paired with radial resolution \zdelta=0.0025, 0.001, 0.0005.  These correspond roughly to a spatial resolution of 6 Mpc, 3 Mpc, and 1.5 Mpc, respectively, in the chosen redshift range.
The angular separation range was from 0 to $\theta_{\rm max}=5^\circ$, divided into 100 bins, chosen as to safely cover the range of interest $r\in[0,200]$ 200 Mpc.
We then ran \realspace\ to convert the RCF into real-space 2PCF and its multipoles, with simulation parameters ($\Omega_m=0.25,h=0.7$).
For this test, shear analysis was turned off.

 \begin{figure}
\includegraphics[width=1\linewidth]{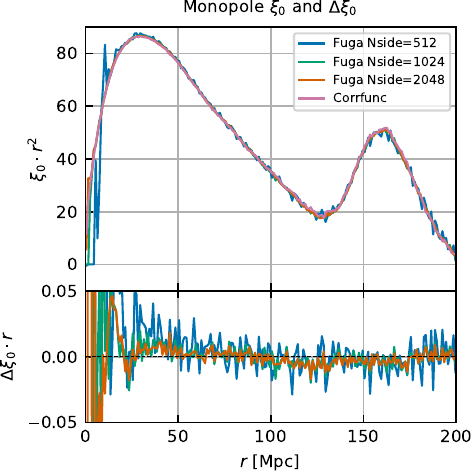}
\caption{2PCF monopole from \fuga, and from exact pair counting (\corrfunc). The input catalog is the full MICE simulation in $z=1.0-1.2$.
\fuga\ results are shown for three resolution settings (see main text for details). The lower panel shows the difference with respect to the \corrfunc\ result.
\label{fig:realcorr0}}
\end{figure}

\begin{figure}
\includegraphics[width=1\linewidth]{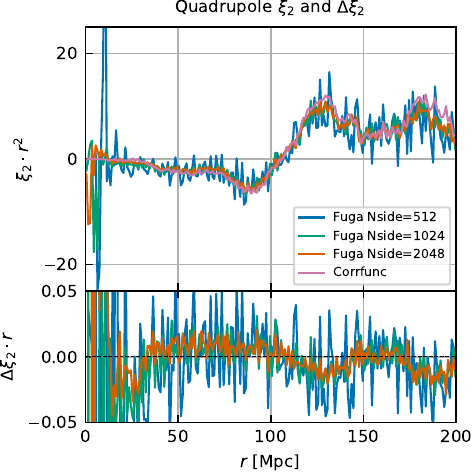}
\caption{Same as Fig.~\ref{fig:realcorr0}, but for the quadrupole.
\label{fig:realcorr2}}
\end{figure}
For comparison, we computed the 2PCF from the same catalog with \corrfunc, which applies the full Landy--Szalay estimator.
We used routines from the \texttt{astropy} library to first convert the redshifts into comoving distance, again assuming the parameters of the simulation.
\corrfunc\ requires an unclustered random catalog, which defines the reference density.  
We generated a random catalog five times the size of the galaxy catalog, $N_{\rm rand}=5\,N_{\rm data}\approx 460\cdot 10^6 $.
The chosen random size was a trade-off between computation time and accuracy.
The redshift sampling was based on one of \fuga's output files \texttt{zweights}, which contains the central redshifts $z_i$ of the redshift bins,
and the number of galaxies in each bin. Bin edges $e_i$ were reconstructed so that each $z_i$ sits at the center of its bin. The pair $(w_i, e_i)$ was passed to \texttt{scipy.stats.rv\_histogram}, which (1) normalizes the weights to obtain a piece-wise constant PDF, (2) forms the corresponding cumulative distribution function \(P_{z}(z)\), and (3) returns $N_{\rm rand}$ inverse-transform samples \(z=P_{z}^{-1}(u)\) with $u\sim\mathcal U(0,1)$.
This yields a redshift distribution that mimics that of the data catalog.
The (RA, DEC) coordinates were pulled from a uniform distribution in the survey area.
Uniform coverage on the sphere corresponds to a PDF that is constant in solid angle
\be
\rm d\Omega = \cos\delta\rm d\delta = \rm d(\sin\delta) \rm d\alpha 
\ee
where $\delta$ is declination and $\alpha$ is right ascension.
To achieve uniform sampling in solid angle, we need to sample $\sin\delta$ uniformly. We did this by drawing a uniform deviate $u\sim \mathcal{U}(0,1)$ and setting $\sin\delta = \sin\delta_{\rm{min}} + u (\sin\delta_{\rm{max}} - \sin\delta_{\rm{min}})$,
where $[\delta_{\rm{min}}, \delta_{\rm{max}}]$ are the declination bounds of the data, in our MICE octant $\delta_{\rm{min}}=0, \delta_{\rm{max}}=\pi/2$.
The right ascension values $\alpha$ were pulled from a uniform distribution in $[\alpha_{\rm min},\alpha_{\rm max}]$.

The \corrfunc\ run took 40 hours on 40 cores, while the \fuga\ runs on the same data took 151, 727, and 2797 seconds, with $N_{\rm side}$=512, 1024, 2048, respectively.
The \corrfunc\ run time was fully dominated by random-random pair counting. 
It is possible to reduce the computational cost of the random pair counts, for instance by splitting the random catalog into small subcatalogs and co-adding the pair counts \citep{Keihanen:2019vst}. In this validation test, the aim was to compare the resulting 2PCF rather than the run times, and we did not pay attention to the optimization of the \corrfunc\ part.

The results of the validation test are shown in Figs. \ref{fig:realcorr0} and \ref{fig:realcorr2}. As expected, there are small differences,
due to the discretization in \fuga, and the finite size of the random catalog in \corrfunc, but overall the results agree well.
The effect of discretization on the \fuga\ correlation function is evident at the smallest distances ($r<10$ Mpc).
The differences between \fuga\ and \corrfunc\ decrease with increasing resolution, but cannot expected to fully disappear at any resolution, because the finite size of the random catalog inevitably causes residual fluctuations in the \corrfunc\ 2PCF, which are not present in the \fuga\ 2PCF. There is also a small residual difference in the quadrupole, likely due to different weighting of data when computing the multipoles (direct integration in \corrfunc, linear fit in \fuga).

\subsection{Power spectrum from discrete coordinates}

In another test, we validated the accuracy of the point-source power spectrum computation described in Sect. \ref{sec:harmonic}. For this purpose, we computed the angular power spectrum for a single redshift shell $z\in[1.0,1.1]$. We did this first using discrete coordinates, which is the default method in \fuga, and then with a map-based SHT. For the latter, we used an older version of \fuga, where this was implemented and maintained as an option for validation purposes. The test was run on a MacBook laptop with 8 cores.

\begin{figure}
    \centering
    \includegraphics[width=1\linewidth]{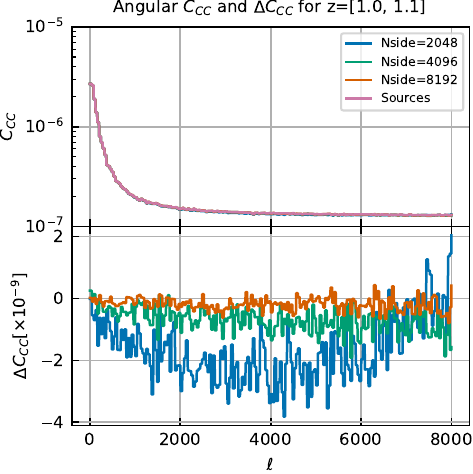}
    \caption{Angular clustering pseudo-spectrum, from exact point-source coordinates and from maps of different resolutions. The lower panel shows the difference with respect to the point-source result.}
    \label{fig:fuga_spectra_CC}
\end{figure}

Figure \ref{fig:fuga_spectra_CC} shows the point-source based clustering spectrum, together with map-based spectra with resolution \nside=2048, 4096, 8192.  The spectra shown are pseudo-spectra, i.e. not corrected for the survey footprint. As the resolution increases, the map-based spectrum approaches that from point source coordinates, validating the point-source algorithm. Computing the point-source spectrum took 22.1 s, while the map-based transform took 10.5 s, 20.3 s, 43.4 s, and 128.1 s with \nside=1024, 2048, 4096, 8192, respectively. Thus, the point-source transform is comparable in speed to the map-based transform with \nside=2048, but reaches a much higher accuracy.

\begin{figure}
    \centering
    \includegraphics[width=1\linewidth]{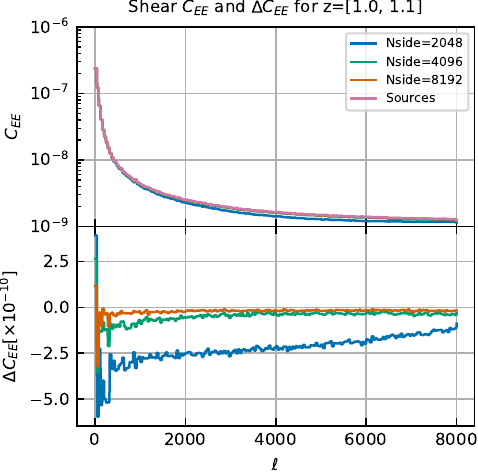}
    \caption{Same as Fig.~\ref{fig:fuga_spectra_CC}, but for $EE$ shear spectrum. The pseudo-spectra have been scaled by the sky fraction to make them comparable (see main text for explanation).}
    \label{fig:fuga_spectra_EE}
\end{figure}

Figure \ref{fig:fuga_spectra_EE} shows a similar comparison for the $EE$ shear spectrum. Here, the uncorrected pseudo-spectra cannot be compared directly, since there is no common normalization. Increasing the resolution further beyond the point where there is only one galaxy per pixel, scales the spectrum down as inversely proportional to the pixel size squared, $\propto 1/N_{\rm side}^4$. To bring the spectra to a common normalization, we scale each spectrum with the sky coverage, which we read from the $\ell=0$ element of the mask spectrum $W^{\rm ss}$.

\subsection{Correlation function from power spectrum}

\begin{figure}
    \centering
    \includegraphics[width=1\linewidth]{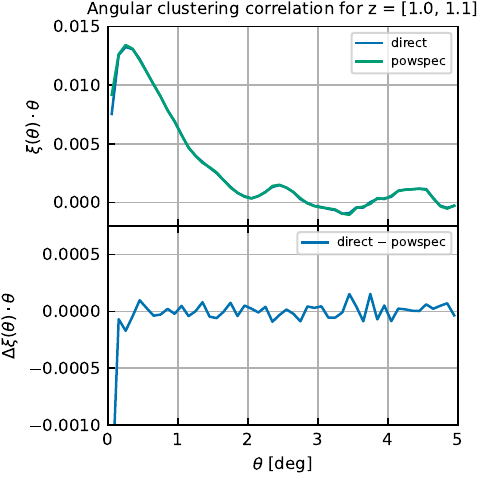}
    \caption{Angular clustering correlation for $z\in[1.0,1.1]$, by direct computation (\nside=2048), or from power spectrum (\lmax=8000). The bottom panel shows the difference.}
    \label{fig:angcorr_GC}
\end{figure}

\begin{figure}
    \centering
    \includegraphics[width=1\linewidth]{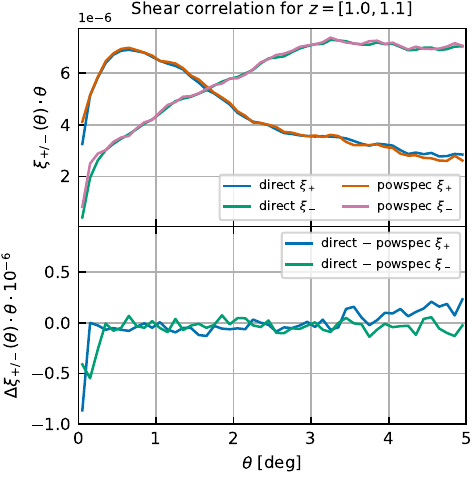}
    \caption{Same as Fig. \ref{fig:angcorr_GC}, but angular shear correlation ($\xi_+,\xi_-$).}
    \label{fig:angcorr_shear}
\end{figure}

In the third validation test, we compared correlation functions computed directly or from power spectra. We selected the redshift range $z\in[1.0,1.1]$ for consideration and computed the angular correlation function in two ways. First, we computed the correlation function through the modified Landy--Szalay, as described in Sections \ref{sec:clusteringCorrelation}--\ref{sec:shearCorrelation}. We refer to this as the ``direct method''. As an alternative method, we (1) computed the angular power spectrum using the exact galaxy coordinates, (2) subtracted an estimate for the shot noise, which we obtained as the mean of the last 500 multipoles, (3) applied the kernel correction as described in Sect. \ref{sec:skymask}, and (4) transformed the power spectrum into a correlation function as per Sect. \ref{sec:powspectheory}. The results are shown in Figs. \ref{fig:angcorr_GC} and \ref{fig:angcorr_shear}. The clustering correlation function from these two different methods agrees very well, further validating our code.

%The direct computation of the correlation function took 157 s with $N_{\rm side}=1024$ resolution. Computing the angular power spectrum to $\ell_{\rm max}=8000$ took 24 s. The construction of the coupling kernel took 98 s, increasing the cost of the power-spectrum method to 122 s. In this test, the power-spectrum approach was thus slightly more efficient. The relative speed of the two methods depends on the parameter settings. The cost of direct pair counting strongly depends on the resolution chosen. For instance, increasing $N_{\rm side}$ from 1024 to 2048 increases the wall-clock time from 157 s to 2037 s. The cost of the power-spectrum approach, on the other hand, increases with increasing $\lmax$.  Going from $\lmax=6000$ to $\lmax=8000$ increases the cost of the kernel correction from 98 s to 235 s. It would be premature to declare one method or the other as the more efficient one.

\subsection{Considerations of speed}

\begin{figure}
    \centering
    \includegraphics[width=1\linewidth]{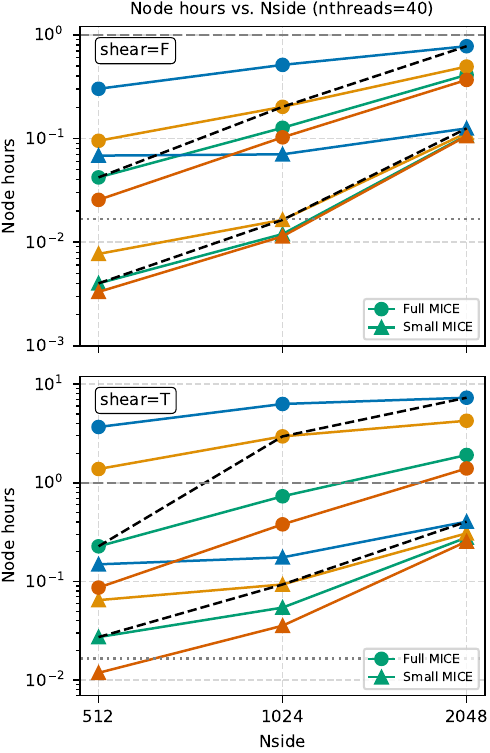}
    \caption{CPU cost of RCF computation, as a function of pixel resolution \nside, and different redshift resolutions \zdelta, for clustering (top), or for clustering and shear (bottom).
    The redshift resolution is, from top down,
    \zdelta=0.0005 (blue), \zdelta=0.001 (yellow), \zdelta=0.0025 (green), zdelta=0.005 (red).
    The dashed lines connect points where \nside\ and \zdelta\ correspond to the same spatial resolution (from left to right, 6 Mpc, 3 Mpc, 1.5 Mpc). 
    The input data is the full (circles) or the small (triangles) MICE simulation in $z=1.0-1.2$. To guide the eye, one node-hour and one node-minute have been indicated by horizontal lines.}
    \label{fig:timings_nside}
\end{figure}

Figure \ref{fig:timings_nside} shows the CPU cost of computing the RCF for the redshift range $z=1.0-1.2$ with \fuga, for different combinations of \nside\ and \zdelta.
All timing tests were run on a single node of the Puhti supercluster. We show separately the CPU cost with or without shear analysis.
The dashed lines indicate 
the cases where the angular and redshift resolutions correspond to the same physical distance:
$(\nside=512,\zdelta=0.0025)$, $(\nside=1024,\zdelta=0.001)$, and
$(\nside=2048,\zdelta=0.0005)$.
In the chosen redshift range these combinations correspond roughly to a spatial resolution of 6 Mpc, 3 Mpc, and 1.5 Mpc, respectively.
Changing both resolutions in unison is meaningful when the goal is to estimate the real-space correlation function. 
In tomographic analysis, one might be more interested in the CPU cost for a fixed redshift resolution. 
Results are shown for small and full MICE inputs, and with and without shear analysis. 

\begin{figure}
    \centering
    \includegraphics[width=1\linewidth]{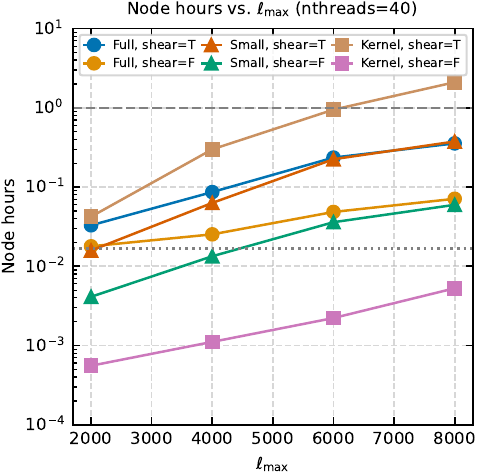}
    \caption{CPU cost of harmonic-space analysis of \fuga, for 630 redshift-shell pairs. Shown is the cost of computing the redshift-space pseudo-spectrum for \lmax=2000--8000 for full (circles) and small (triangles) MICE in $z=1.0-1.2$.  The redshift resolution was fixed at \zdelta=0.005. The squares indicate the cost of constructing the mode-mode coupling kernel to correct for the survey footprint.}
    \label{fig:timings_lmax}
\end{figure}

We further study the CPU cost of the harmonic analysis. In Fig.~\ref{fig:timings_lmax} we show the CPU cost of computing the redshift-space power spectrum as a function of the $\lmax$ parameter. In this test, we fix the redshift resolution to \zdelta=0.005 and compute the auto- and cross-spectra for $|z_1-z_2|<0.1$ in the range $z\in 1.0-1.2$. This involves computing the SHT for 40 redshift bins, and constructing spectra for 630 pairs of bins.
We show the cost of computing the raw pseudo-spectrum for the small and the full MICE catalogs, with and without shear analysis.
In the same graph, we also show the cost of the post-processing step that corrects for the survey mask.
The cost of full power-spectrum analysis is the sum of the cost of pseudo-spectrum estimation and that of post-processing. 

The post-processing cost is dominated by the construction of the coupling kernel and inverting it.
The kernel correction does not depend on the catalog size; thus the cost is the same for the small or the full catalog.
As explained in Sect. \ref{sec:skymask}, a single coupling kernel applies to all clustering spectra,
while for shear, each of the 630 redshift pairs has its own kernel which depends on the distribution of the galaxies. For this reason, the kernel correction is orders of magnitude more expensive in the shear analysis.

\section{Conclusions}

We have demonstrated the usefulness of the redshift-space correlation function (RCF) in the estimation of the real-space two-point correlation function. Once the redshift-space correlation function has been constructed and stored, it is very fast to compute the real-space two-point correlation function (2PCF) for any given cosmology.   

We have developed the \fuga\ code that constructs the RCF, and its harmonic counterpart, the redshift-space power spectrum (RP). The direct RCF computation is based on a modified Landy-Szalay estimator, and relies on a discretized spherical grid, and the assumption that the selection function is piecewise isotropic in the survey light-cone. The code is very fast: Computing the RCF and the two-point correlation function for the full MICE simulation in $z=1.0-1.2$ at resolution settings that correspond to 1.5 Mpc spatial resolution (nside=2048, \zdelta=0.0005), took 47 node-min for clustering only, and 7.3 node-hours with shear analysis included.
Relaxing the resolution to 3 Mpc drops the times to 12 node-min for clustering, and to 3 node-hours for clustering and shear.

\fuga\ estimates power spectra using the exact coordinates of the galaxies, and thus requires discretization only in the radial dimension.
The power spectrum can be converted to a correlation function, with the help of Wigner functions, providing an alternative way of constructing the RCF.  The question of which of the two methods is the most efficient cannot be answered definitively, since different parts of the code scale differently, and the optimal parameter settings depend on the characteristics of the data set and on the accuracy requirements.
Both methods have their benefits and drawbacks.
The direct method requires less run-time memory than the power-spectrum procedure. The latter has a large memory footprint, because a large number of SHTs must be kept in memory simultaneously to construct the cross-spectra between $(z_k,z_{k'})$ bins.  The power-spectrum approach, on the other hand, offers the benefit of accommodating an arbitrary selection function, while the direct method relies on a piecewise isotropic selection function.

An important future development is to test the use of RCF in the context of cosmological parameter estimation. The RCF approach provides the great benefit that it does not require a fiducial cosmological model.  The RCF can be used in a parameter sampling loop in two different ways:

1. Precompute the RCF once for the dataset. At every sampling step, convert the RCF to a real-space correlation function using the current parameter values and evaluate the likelihood at the 2PCF level. This ensures that the 2PCF is always consistent with the sampling parameters. The data vector for a likelihood in this case is the 2PCF.

2. Use the precomputed RCF itself as the dataset. At each sampling step, construct a model for the RCF using the current parameters and evaluate the likelihood at the RCF level. The data vector in this case is the RCF.

Approach 1 is more straightforward, as it can largely be realized using existing software tools, with relatively little extra work. Approach 2 involves the more demanding step of modeling the RCF from a given cosmology, or from a matter power spectrum provided by a standard software tool, such as CAMB. Approach 2, on the other hand, is conceptually simpler in the sense that it does not involve modifying the data between sampling steps.

Parameter sampling requires an estimate for the covariance of the data.  \fuga\ provides a diagonal covariance estimate that is based on number counts, and thus captures the Poisson part of the full covariance.  Because \fuga\ is very fast, a full covariance could be constructed as a sample covariance from a number of mock catalogs, possibly accompanied by some smoothing or regularization step, but this remains to be tested. An interesting question is whether the RCF can be compressed in some way without losing information. All this is a field of study on its own right and is left for future work.

\fuga\ is available via: \\
\url{https://wiki.helsinki.fi/xwiki/bin/view/FuGa3D/}

\begin{acknowledgements}
This work has been supported by Research Council of Finland grant 347088.
    The authors wish to acknowledge CSC – IT Center for Science, Finland, for computational resources.
\end{acknowledgements}

\bibliographystyle{aa} % style aa.bst
\bibliography{fuga_bibliography.bib}

\include{appendix.tex}

\end{document}

%% file: appendix.tex
\onecolumn

\begin{appendix}

\section{From power spectrum to correlation function}
\label{appendix:xi}

Here we derive the relation between the angular correlation function and power spectrum.
We expand the desnity contrast and the shear field in spin-spherical harmonics:
\bea
\delta (\vartheta,\varphi) &=& \sum_{\ell m} a_{\ell m}\, {}_0Y_{\ell m}(\vartheta,\varphi) \nonumber \\
(\gamma_1\pm i\gamma_2) (\vartheta,\varphi) &=& \sum_{\ell m}\,{}_{\pm2}a_{\ell m}\, {}_{\pm2}Y_{\ell m} (\vartheta,\varphi)
= -(a_{E\ell m}\pm ia_{B\ell m})\, {}_{\pm2}Y_{\ell m}(\vartheta,\varphi) 
\eea
The $\gamma_1,\gamma_2$ components  can be solved from this as
\bea
 \gamma_1 (\vartheta,\varphi) &=& -\tfrac12\sum_{\ell m} a_{E\ell m} ({}_2Y_{\ell m}(\vartheta,\varphi)+{}_{-2}Y_{\ell m}(\vartheta,\varphi)) 
 -\tfrac i2\sum_{\ell m} a_{B\ell m} ({}_2Y_{\ell m}(\vartheta,\varphi)-{}_{-2}Y_{\ell m}(\vartheta,\varphi)) \nonumber \\
 \gamma_2 (\vartheta,\varphi) &=& -\tfrac12\sum_{\ell m} a_{B\ell m} ({}_2Y_{\ell m}(\vartheta,\varphi)+{}_{-2}Y_{\ell m}(\vartheta,\varphi)) 
 +\tfrac i2\sum_{\ell m} a_{E\ell m} ({}_2Y_{\ell m}(\vartheta,\varphi)-{}_{-2}Y_{\ell m}(\vartheta,\varphi)) 
 \eea
 The clustering correlation function is
 \be
 \xi(\theta) = \langle\delta^\ast(\vartheta,\varphi) \delta(\vartheta',\varphi')  \rangle 
 = \sum_{\ell m}\sum_{\ell'm'} \langle a^\ast_{\ell m}a_{\ell' m'}\rangle \, {}_0Y^\ast_{\ell m}(\vartheta,\varphi)\, {}_0Y_{\ell' m'}(\vartheta',\varphi') 
 = \sum_{\ell} P_\ell \sum_m{}_0Y^\ast_{\ell m}(\vartheta,\varphi) \,{}_0Y_{\ell m}(\vartheta',\varphi') 
 \ee
 where $\theta$ is the angular separation between points $(\vartheta,\varphi)$ and $(\vartheta',\varphi')$.
Other correlation components involve similar combinations of two spin harmonics.
We introduce the short-hand notation $S_{\ell,s_1,s_2}$ and write, using
the general addition theorem of spin harmonics,
\be
S_{\ell,s_1,s_2} \equiv \sum_m \,{}_{s_1}Y^\ast_{\ell m}(\vartheta,\varphi) \,{}_{s_2}Y_{\ell m}(\vartheta',\varphi')
= \sqrt{\frac{2\ell+1}{4\pi}}\,{}_{s_2}Y_{\ell,-s_1}(\theta,\phi)e^{-is_2\psi}
= \sqrt{\frac{2\ell+1}{4\pi}}\,d^{\ell}_{-s_1,-s_2}(\theta)e^{-is_1\phi}e^{-is_2\psi}
\ee
where $\phi,\theta,\psi$ are the Euler angles that rotate $(\vartheta,\varphi)$ to $(\vartheta',\varphi')$.
In the last equality we expressed the spherical harmonic in terms of the reduced Wigner function.
From the symmtery properties of Wigner functions it follows $S_{\ell,2,2}=S_{\ell,-2,-2}$, $S_{\ell,2,-2}=S_{\ell,-2,2}$,
and $S_{\ell,0,2}=S_{\ell,2,0}=S_{\ell,-2,0}=S_{\ell,0,-2}$.

Without loss of generality we can place the pair of points on the same meridian, so that $\phi=\psi=0$,
and $\theta$ remains as the angular separation between the points.
We can then interpret $\gamma_{\rm t}=\gamma_1$ and $\gamma_{\times}=\gamma_2$.
The independent components are then
\be
S_{\ell,0,0}=\sqrt{\frac{2\ell+1}{4\pi}}d^\ell_{0,0}(\theta), \qquad
S_{\ell,2,2}=\sqrt{\frac{2\ell+1}{4\pi}}d^\ell_{2,0}(\theta), \qquad
S_{\ell,2,-2}=\sqrt{\frac{2\ell+1}{4\pi}}d^\ell_{2,-2}(\theta)
\ee
We can now construct the correlators
\bea
\xi_{\rm cc}(\theta) &=& \langle \delta^\ast\delta\rangle 
 = P^{CC}_\ell S_{\ell,0,0} \nonumber \\
\xi_{\rm ct}(\theta) &=& \langle \delta^\ast\gamma_1\rangle =
 -\tfrac12 P^{CE}_\ell(S_{\ell,0,2}+S_{\ell,0,-2})  -\tfrac i2 P^{CB}_\ell(S_{\ell,0,2}-S_{\ell,0,-2})  
 = -P^{CE}_\ell S_{\ell,0,2} \nonumber \\
 \xi_{\rm c\times}(\theta) &=& \langle \delta^\ast\gamma_2\rangle =
 -\tfrac12 P^{CB}_\ell(S_{\ell,0,2}+S_{\ell,0,-2})  +\tfrac i2 P^{CE}_\ell(S_{\ell,0,2}-S_{\ell,0,-2})  
 = -P^{CB}_\ell S_{\ell,0,2} \nonumber \\
 \xi_{\rm tt}(\theta) &=& \langle \gamma_1^\ast\gamma_1\rangle =
 \tfrac14 P^{EE}_\ell(S_{\ell,2,2}+S_{\ell,2,-2}+S_{\ell,-2,2}+S_{\ell,-2,-2})  
 +\tfrac14 P^{BB}_\ell(S_{\ell,2,2} -S_{\ell,2,-2}-S_{\ell,-2,2} +S_{\ell,-2,-2})  \nonumber \\
 &&\qquad\quad\ +\tfrac i4 P^{EB}_\ell(S_{\ell,2,2}-S_{\ell,2,-2}+S_{\ell,-2,2}-S_{\ell,-2,-2}))  
 +\tfrac i4 P^{BE}_\ell(-S_{\ell,2,2} -S_{\ell,2,-2} +S_{\ell,-2,2} +S_{\ell,-2,-2}))   \nonumber \\
 && \qquad \ = P^{EE}_\ell(S_{\ell,2,2}+S_{\ell,2,-2}) +P^{BB}_\ell (S_{\ell,2,2}-S_{\ell,2,-2}) \nonumber \\
 \xi_{\times\times}(\theta) &=& \langle \gamma_2\gamma_2\rangle = 
\tfrac14 P^{BB}_\ell(S_{\ell,2,2}+S_{\ell,2,-2}+S_{\ell,-2,2}+S_{\ell,-2,-2})  
 +\tfrac14 P^{EE}_\ell(S_{\ell,2,2}-S_{\ell,2,-2}-S_{\ell,-2,2} +S_{\ell,-2,-2})  \nonumber \\
 &&\qquad \quad = P^{BB}_\ell(S_{\ell,2,2}+S_{\ell,2,-2}) +P^{EE}_\ell (S_{\ell,2,2}-S_{\ell,2,-2}) \nonumber \\
 \xi_{\rm t\times}(\theta) &=& \langle \gamma_1\gamma_2\rangle = 
 \tfrac14 P^{EB}_\ell(S_{\ell,2,2}+S_{\ell,2,-2}+S_{\ell,-2,2}+S_{\ell,-2,-2})  
 +\tfrac14 P^{BE}_\ell(-S_{\ell,2,2}+S_{\ell,2,-2}+S_{\ell,-2,2}-S_{\ell,-2,-2})  \nonumber \\
 && \qquad\quad = P^{EB}_\ell(S_{\ell,2,2}+S_{\ell,2,-2}) +P^{BE}_\ell (-S_{\ell,2,2}+S_{\ell,2,-2}) \nonumber \\
 \xi_{\times\rm t}(\theta) &=& \langle \gamma_2\gamma_1\rangle = 
 \tfrac14 P^{BE}_\ell(S_{\ell,2,2}+S_{\ell,2,-2}+S_{\ell,-2,2}+S_{\ell,-2,-2})  
 +\tfrac14 P^{EB}_\ell(-S_{\ell,2,2}+S_{\ell,2,-2}+S_{\ell,-2,2}-S_{\ell,-2,-2}))  \nonumber \\
 && \qquad\quad = P^{BE}_\ell(S_{\ell,2,2}+S_{\ell,2,-2}) +P^{EB}_\ell (-S_{\ell,2,2}+S_{\ell,2,-2}) 
\eea
After the line for $\xi_{\rm tt}$ we did not explicitly write the imaginary elements, which vanish anyway.
The correlated fields here may represent two different fields, for instance from two redshift shells. For this reason the ordering of the elements on the right-hand-side is relevant, and in general $P_\ell^{EB}\neq P_\ell^{BE}$. 

The correlations can be combined into independent combinations as
\bea
\xi_{\rm cc}(\theta) &=& \sum_\ell \frac{2\ell+1}{4\pi} C^{CC}_\ell d^\ell_{00}(\theta) \nonumber \\
\xi_+(\theta) &=& \sum_\ell \frac{2\ell+1}{4\pi} (C^{EE}_\ell +C^{BB}_\ell ) d^\ell_{2,2}(\theta) \nonumber \\
\xi_-(\theta) &=& \sum_\ell \frac{2\ell+1}{4\pi} (C^{EE}_\ell -C^{BB}_\ell) d^\ell_{2,-2}(\theta) \nonumber \\
\xi_{\rm ct}(\theta) &=& -\sum_\ell \frac{2\ell+1}{4\pi} C^{CE}_\ell  d^\ell_{20}(\theta) \nonumber \\
\xi_{\rm c\times}(\theta) &=& -\sum_\ell \frac{2\ell+1}{4\pi} C^{CB}_\ell  d^\ell_{20}(\theta) \nonumber \\
\xi_{\rm t\times}(\theta) +\xi_{\times\rm t}(\theta) &=& 
   \sum_\ell \frac{2\ell+1}{4\pi} (C^{EB}_\ell +C^{BE}_\ell )  d^\ell_{2,-2}(\theta) \nonumber \\
\xi_{\rm t\times}(\theta) -\xi_{\times\rm t}(\theta) &=& 
  \sum_\ell \frac{2\ell+1}{4\pi} (C^{EB}_\ell -C^{BE}_\ell )  d^\ell_{2,2}(\theta) 
\eea
Note that $d^\ell_{00}(\theta)=L_\ell(\cos\theta)$, where $L_\ell$ is the Legendre polynomial.

%%%%%%%%%%%%%%%%%
\section{Coupling kernel}
\label{appendix:kernel}

The density field observed under a binary sky mask $w^{\rm c}(\vartheta,\varphi)=1,0$ is
\be
\hat \delta(\vartheta,\varphi) = \delta(\vartheta,\varphi) w^{\rm c}(\vartheta,\varphi) 
=\sum_{\ell'm'} \sum_{\ell''m''} a_{\ell'm'} \omega_{\ell''m''} Y_{\ell'm'}(\vartheta,\varphi)  Y_{\ell''m''}(\vartheta,\varphi) 
\ee
Similarly, for shear,
\be
(\hat \gamma_1\pm i\hat\gamma_2)(\vartheta,\varphi) = (\gamma_1+i\gamma_2)(\vartheta,\varphi) w^{\rm s}(\vartheta,\varphi) 
=\sum_{\ell'm'} \sum_{\ell''m''} {}_{\pm2}a_{\ell'm'} \omega^{\rm s}_{\ell''m''} {}_{\pm2}Y_{\ell'm'}(\vartheta,\varphi)  Y_{\ell''m''}(\vartheta,\varphi) 
\ee
The SHT of the masked field yields
\be
\hat a_{\ell m} = \int \hat \delta(\vartheta,\varphi) Y^\ast_{\ell m}(\vartheta,\varphi) d\Omega
= \sum_{\ell'm'} \sum_{\ell''m''} {}_{}a_{\ell'm'} \omega^{\rm c}_{\ell''m''} \int  Y^\ast_{\ell m}(\vartheta,\varphi) Y_{\ell'm'}(\vartheta,\varphi)  Y_{\ell''m''}(\vartheta,\varphi) d\Omega
\ee
and
\be
{}_{\pm2}\hat a_{\ell m} = \int (\hat \gamma_1\pm\hat\gamma_2) (\vartheta,\varphi) {}_{\pm2}Y^\ast_{\ell m}(\vartheta,\varphi) d\Omega
= \sum_{\ell'm'} \sum_{\ell''m''} {}_{\pm 2}a_{\ell'm'} \omega^{\rm s}_{\ell''m''} \int  {}_{\pm2}Y^\ast_{\ell m}(\vartheta,\varphi) {}_{\pm2}Y_{\ell'm'}(\vartheta,\varphi)  Y_{\ell''m''}(\vartheta,\varphi) d\Omega
\ee
Using the symmetry property ${}_sY^\ast_{\ell m}(\vartheta,\varphi) =(-1)^{m+s}{}_{-s}Y_{\ell -m}(\vartheta,\varphi) $
and the expression of the triple integral in terms of Wigner 3j symbols, 
\be
\int {}_{s}Y^\ast_{\ell m}(\vartheta,\varphi) {}_{s'}Y_{\ell'm'}(\vartheta,\varphi)  {}_{s''}Y_{\ell''m''}(\vartheta,\varphi) d\Omega
=F(\ell,\ell',\ell'') 
\begin{pmatrix} \ell & \ell' & \ell'' \\ m & m' & m'' \end{pmatrix}
\begin{pmatrix} \ell & \ell' & \ell'' \\ -s & -s' & -s'' \end{pmatrix}
\ee
where
\be
F(\ell,\ell',\ell'') = \left[ \frac{(2\ell+1)(2\ell'+1)(2\ell''+1)}{4\pi} \right]^{1/2} 
\ee
we arrive at
\be
\hat a_{\ell m} = \sum_{\ell'm'} \sum_{\ell''m''} \omega^{\rm c}_{\ell''m''} (-1)^m
F(\ell,\ell',\ell'') 
\begin{pmatrix} \ell & \ell' & \ell'' \\ 0 & 0 & 0 \end{pmatrix}
\begin{pmatrix} \ell & \ell' & \ell'' \\ -m & m' & m'' \end{pmatrix} a_{\ell'm'} 
\ee
and
\be
{}_{\pm2}\hat a_{\ell m} = \sum_{\ell'm'} \sum_{\ell''m''}\omega^{\rm s}_{\ell''m''} (-1)^m
F(\ell,\ell',\ell'') 
\begin{pmatrix} \ell & \ell' & \ell'' \\ \pm2 & \mp2 & 0 \end{pmatrix}
\begin{pmatrix} \ell & \ell' & \ell'' \\ -m & m' & m'' \end{pmatrix}  {}_{\pm2}a_{\ell'm'} 
\ee
Further with $\hat a_{E\ell m}=-\tfrac12({}_2\hat a_{\ell m}+{}_{-2}\hat a_{\ell m})$, $\hat a_{B\ell m}=\tfrac i2({}_2\hat a_{\ell m}-{}_{-2}\hat a_{\ell m})$ 
and ${}_{\pm2}a_{\ell m}=-(a_{E\ell m}\pm ia_{B\ell m})$, and the symmetry relation
\be
\begin{pmatrix} \ell & \ell' & \ell'' \\ -2 & 2 & 0 \end{pmatrix} = (-1)^{\ell+\ell'+\ell''}
\begin{pmatrix} \ell & \ell' & \ell'' \\ 2 & -2 & 0 \end{pmatrix} 
\ee
we find
\bea
\hat a_{E\ell m} &=& \sum_{\ell'm'} \sum_{\ell''m''} \omega^{\rm s}_{\ell''m''} (-1)^m
F(\ell,\ell',\ell'') 
\begin{pmatrix} \ell & \ell' & \ell'' \\ 2 & -2 & 0 \end{pmatrix}
\begin{pmatrix} \ell & \ell' & \ell'' \\ -m & m' & m'' \end{pmatrix}
\left( \tfrac12 a_{E\ell' m'} (1+ (-1)^{\ell+\ell'+\ell''}) + \tfrac i2 a_{B\ell' m'} (1- (-1)^{\ell+\ell'+\ell''}) \right) \nonumber \\
\hat a_{B\ell m} &=& \sum_{\ell'm'} \sum_{\ell''m''} \omega^{\rm s}_{\ell''m''} (-1)^m
F(\ell,\ell',\ell'') 
\begin{pmatrix} \ell & \ell' & \ell'' \\ 2 & -2 & 0 \end{pmatrix}
\begin{pmatrix} \ell & \ell' & \ell'' \\ -m & m' & m'' \end{pmatrix}
\left( \tfrac12 a_{B\ell' m'} (1+ (-1)^{\ell+\ell'+\ell''}) - \tfrac i2 a_{E\ell' m'} (1- (-1)^{\ell+\ell'+\ell''}) \right)
\eea
The pseudo-spectrum for clustering is
\be
\hat P_\ell =\frac{1}{2\ell+1}\sum_m \hat a^\ast_{\ell m} \hat a_{\ell m}
\ee
Inserting the expression for $\hat a_{\ell m}$ and arranging the sums we find
\bea
\langle \hat P_\ell \rangle &=& 
\frac{1}{2\ell+1} \sum_{\ell'\ell''\ell_1\ell_2}
F(\ell,\ell',\ell'') F(\ell,\ell_1,\ell_2) 
\begin{pmatrix} \ell & \ell' & \ell'' \\ 0 & 0 & 0 \end{pmatrix}
\begin{pmatrix} \ell & \ell_1 & \ell_2 \\ 0 & 0 & 0 \end{pmatrix} \nonumber \\
&&\qquad\times
\sum_{mm'm''m_1m_2}
\omega^{\rm c}_{\ell''m''}  \omega^{\rm c}_{\ell_2 m_2}   
\begin{pmatrix} \ell & \ell' & \ell'' \\ -m & m' & m'' \end{pmatrix} 
\begin{pmatrix} \ell & \ell_1 & \ell_2 \\ -m & m_1 & m_2 \end{pmatrix} 
\langle a^\ast_{\ell' m'} a_{\ell_1 m_1} \rangle 
\eea
Using first
\be
\langle a^\ast_{\ell' m'} a_{\ell_1 m_1} \rangle = \delta_{\ell'\ell_1}\delta_{m' m_1} P_\ell
\ee
and then
\be
\sum_{mm'}
\begin{pmatrix} \ell & \ell' & \ell'' \\ -m & m' & m'' \end{pmatrix} 
\begin{pmatrix} \ell & \ell_1 & \ell_2 \\ -m & m_1 & m_2 \end{pmatrix} 
=\frac1{2\ell''+1} \delta_{\ell''\ell_2} \delta_{m''m_2}
\ee
and finally writing for the mask spectrum
\be
\sum_{m''} \omega^{\rm c\ast}_{\ell''m''} \omega^{\rm c}_{\ell''m''} = (2\ell''+1)W^{\rm cc}_{\ell''}
\ee
we arrive at the final result
\be
\langle \hat P_\ell \rangle = \sum_{\ell'}P_{\ell}
\sum_{\ell''} W^{\rm cc}_{\ell''} \frac{1}{4\pi}(2\ell'+1)(2\ell''+1) \begin{pmatrix} \ell & \ell' & \ell'' \\ 0 & 0 & 0 \end{pmatrix}^2
\ee
From here one can read off the coupling kernel for clustering.

The other pseudo-spectra can be worked on the same way. The reduction of the multiple sums happens in the same way, but we get a different combination of spectra on the right-hand-side. Specifically, for the cross-spectra
\bea
\langle \hat P_\ell^{CE}\rangle &\propto& \left\langle a^\ast_{\ell' m'} \left( \tfrac12 a_{E\ell' m'} (1+ (-1)^{\ell+\ell'+\ell''}) + \tfrac i2 a_{B\ell' m'} (1- (-1)^{\ell+\ell'+\ell''}) \right) \right\rangle 
= \tfrac12 \left( P_{\ell'}^{CE} (1+ (-1)^{\ell+\ell'+\ell''})  +i P_{\ell'}^{CB}(1- (-1)^{\ell+\ell'+\ell''})   \right)\nonumber \\
\langle \hat P_\ell^{CB} \rangle &\propto& \left\langle a^\ast_{\ell' m'} \left( \tfrac12 a_{B\ell' m'} (1+ (-1)^{\ell+\ell'+\ell''}) - \tfrac i2 a_{E\ell' m'} (1- (-1)^{\ell+\ell'+\ell''}) \right) \right\rangle
=  \tfrac12 \left( P_{\ell'}^{CB} (1+ (-1)^{\ell+\ell'+\ell''})  -i P_{\ell'}^{CE}(1- (-1)^{\ell+\ell'+\ell''}) \right)
\eea
The full expressions for $\langle \hat P_\ell^{CE}\rangle,\langle \hat P_\ell^{CB}\rangle$ include the Wigner symbol $\begin{pmatrix} \ell & \ell' & \ell'' \\ 0 & 0 & 0 \end{pmatrix}$ which vanishes for odd $(\ell+\ell'+\ell'')$. We can assume $(-1)^{ \ell+\ell'+\ell''}=1$ and we have simply $\langle \hat P_\ell^{CE}\rangle \propto P_{\ell'}^{CE}$, $\langle \hat P_\ell^{CB}\rangle \propto P_{\ell'}^{CB}$.

When working on the shear spectra, we must keep in mind that we are considering cross-spectra between two redshift shells. Thus the ordering of the elements is relevant, and $a^\ast_{E\ell m}a_{B\ell m}\neq a_{B\ell m}a^\ast_{E\ell m}$.
\bea
\langle \hat P_\ell^{EE}\rangle &\propto& 
\left\langle 
   \left( \tfrac12 a^\ast_{E\ell' m'} (1+ (-1)^{\ell+\ell'+\ell''}) - \tfrac i2 a^\ast_{B\ell' m'} (1- (-1)^{\ell+\ell'+\ell''}) \right)   
   \left( \tfrac12 a_{E\ell' m'} (1+ (-1)^{\ell+\ell'+\ell''}) + \tfrac i2 a_{B\ell' m'} (1- (-1)^{\ell+\ell'+\ell''}) \right) 
 \right\rangle  \nonumber \\
 && = \tfrac12 P_{\ell'}^{EE}(1+ (-1)^{\ell+\ell'+\ell''})  + \tfrac12P_{\ell'}^{BB}(1- (-1)^{\ell+\ell'+\ell''}) 
\eea
where we used the fact that the factors $\tfrac12(1\pm(-1)^{\ell+\ell'+\ell''})$ are either 0 or 1, thus $[\tfrac12(1\pm(-1)^{\ell+\ell'+\ell''})]^2=\tfrac12(1\pm(-1)^{\ell+\ell'+\ell''})$
and $(1+(-1)^{\ell+\ell'+\ell''})(1-(-1)^{\ell+\ell'+\ell''})=0$.
Similar calculation for the other shear components gives
\bea
\langle \hat P_\ell^{BB}\rangle &\propto&  
%\left\langle 
%   \left( \tfrac12 a^\ast_{B\ell' m} (1+ (-1)^{\ell+\ell'+\ell''}) + \tfrac i2 a^\ast_{E\ell'm} (1- (-1)^{\ell+\ell'+\ell''}) \right)  
%   \left( \tfrac12 a_{B\ell' m} (1+ (-1)^{\ell+\ell'+\ell''}) - \tfrac i2 a_{E\ell'm} (1- (-1)^{\ell+\ell'+\ell''}) \right)
%\right\rangle  \nonumber \\
%&& = 
\tfrac12P_{\ell'}^{BB}(1+ (-1)^{\ell+\ell'+\ell''})  + \tfrac12P_{\ell'}^{EE}(1- (-1)^{\ell+\ell'+\ell''})  \nonumber \\
\langle \hat P_\ell^{EB}\rangle &\propto&  
%\left\langle 
 %  \left( \tfrac12 a^\ast_{E\ell' m} (1+ (-1)^{\ell+\ell'+\ell''}) - \tfrac i2 a^\ast_{B\ell'm} (1- (-1)^{\ell+\ell'+\ell''}) \right)  
%   \left( \tfrac12 a_{B\ell' m} (1+ (-1)^{\ell+\ell'+\ell''}) - \tfrac i2 a_{E\ell'm} (1- (-1)^{\ell+\ell'+\ell''}) \right)
%\right\rangle  \nonumber \\
%&& =  
\tfrac12P_{\ell'}^{EB}(1+ (-1)^{\ell+\ell'+\ell''})  - \tfrac12P_{\ell'}^{BE}(1- (-1)^{\ell+\ell'+\ell''})  \nonumber \\
\langle \hat P_\ell^{BE}\rangle &\propto&  
\tfrac12P_{\ell'}^{BE}(1+ (-1)^{\ell+\ell'+\ell''})  - \tfrac12P_{\ell'}^{EB}(1- (-1)^{\ell+\ell'+\ell''})  
\eea
The results can be arranged into natural components 
\bea
\langle \hat P_\ell^{EE}+\hat P_\ell^{BB} \rangle &\propto&  P_{\ell'}^{EE}+P_{\ell'}^{BB} \nonumber \\
\langle \hat P_\ell^{EE}-\hat P_\ell^{BB} \rangle &\propto&  (P_{\ell'}^{EE}-P_{\ell'}^{BB})(-1)^{\ell+\ell'+\ell''}  \nonumber \\
\langle \hat P_\ell^{EB}+\hat P_\ell^{BE} \rangle &\propto&  (P_{\ell'}^{EB}+P_{\ell'}^{BE})(-1)^{\ell+\ell'+\ell''}  \nonumber \\
\langle \hat P_\ell^{EB}-\hat P_\ell^{BE} \rangle &\propto&  (P_{\ell'}^{EB}-P_{\ell'}^{BE}) 
\eea
Collecting the full results we arrive at
\bea
\langle\hat C^{CC}_\ell \rangle &=& \sum_{\ell'} K^0_{\ell\ell'} C^{CC}_{\ell'} \nn
\langle\hat C^{CE}_\ell \rangle &=& \sum_{\ell'} K^\times_{\ell\ell'} C^{CE}_{\ell'} \nn
\langle\hat C^{CB}_\ell \rangle &=& \sum_{\ell'} K^\times_{\ell\ell'} C^{CB}_{\ell'} \nn
\langle\hat C^{EE}_\ell +\hat C^{BB}_\ell \rangle &=& \sum_{\ell'} K^+_{\ell\ell'} (C^{EE}_{\ell'}+C^{BB}_{\ell'} ) \nn
\langle\hat C^{EE}_\ell -\hat C^{BB}_\ell \rangle &=& \sum_{\ell'} K^-_{\ell\ell'} (C^{EE}_{\ell'}-C^{BB}_{\ell'} ) \nn
\langle\hat C^{EB}_\ell +\hat C^{BE}_\ell \rangle &=& \sum_{\ell'} K^-_{\ell\ell'} (C^{EB}_{\ell'}+C^{BE}_{\ell'} ) \nn
\langle\hat C^{EB}_\ell -\hat C^{BE}_\ell \rangle &=& \sum_{\ell'} K^+_{\ell\ell'} (C^{EB}_{\ell'}-C^{BE}_{\ell'} ) \, .
\eea
where the four types of kernel matrix are
\bea
K^0_{\ell\ell'} &=& \frac{ (2\ell'+1)}{4\pi}\sum_{\ell''} W^{\rm cc}_{\ell''}(2\ell''+1)
   \begin{pmatrix} \ell & \ell' & \ell'' \\ 0 & 0 & 0 \end{pmatrix}^2 \nn
K^\times_{\ell\ell'} &=& \frac{ (2\ell'+1)}{4\pi}\sum_{\ell''} W^{\rm cs}_{\ell''} (2\ell''+1)
   \begin{pmatrix} \ell & \ell' & \ell'' \\ 2 & -2 & 0 \end{pmatrix} 
   \begin{pmatrix} \ell & \ell' & \ell'' \\ 0 & 0 & 0 \end{pmatrix} \nn
 K^-_{\ell\ell'} &=& \frac{ (2\ell'+1)}{4\pi}\sum_{\ell''} W^{\rm ss}_{\ell''} (2\ell''+1)
   \begin{pmatrix} \ell & \ell' & \ell'' \\ 2 & -2 & 0 \end{pmatrix}^2 (-1)^{\ell+\ell'+\ell''} \nn
   K^+_{\ell\ell'} &=& \frac{ (2\ell'+1)}{4\pi}\sum_{\ell''} W^{\rm ss}_{\ell''} (2\ell''+1)
   \begin{pmatrix} \ell & \ell' & \ell'' \\ 2 & -2 & 0 \end{pmatrix}^2
\eea

%%%%%%%%%%%%%%%%%%%%%%

\end{appendix}